\documentclass[twocolumn]{aastex7}

\usepackage{amsmath,amssymb,amsfonts}
\usepackage{xspace}
\usepackage{microtype}
\usepackage{graphicx}

\newcommand{\jwst}{\rm{JWST}\xspace}
\newcommand{\hst}{\rm{HST}\xspace}
\newcommand{\spitzer}{\rm{Spitzer}\xspace}
\newcommand{\paa}{Pa$\alpha$\xspace}

\newcommand{\bra}{Br$\alpha$\xspace}
\newcommand{\brb}{Br$\beta$\xspace}
\newcommand{\arii}{[\ion{Ar}{2}]\xspace}
\newcommand{\ariii}{[\ion{Ar}{3}]\xspace}
\newcommand{\siv}{[\ion{S}{4}]\xspace}
\newcommand{\neii}{[\ion{Ne}{2}]\xspace}

\newcommand{\htwo}{H$_2$\xspace}
\newcommand{\um}{\,\micron\xspace}
\newcommand{\dnps}{DN\,s$^{-1}$\xspace}
\newcommand{\congress}{CONGRESS\xspace}
\newcommand{\fresco}{FRESCO\xspace}
\newcommand{\sapphires}{SAPPHIRES\xspace}
\newcommand{\hdstar}{HD\,163466\xspace}
\newcommand{\lmcpn}{LHA\,120-N\,133\xspace}
\newcommand{\relname}{\texttt{MIRI\_WFSS\_CAL\_v1.2}\xspace}

\newcommand{\Nallexp}{1{,}611\xspace}     
\newcommand{\Nallprog}{113\xspace}        
\newcommand{\Natlas}{180\xspace}          
\newcommand{\NatlasGN}{149\xspace}
\newcommand{\NatlasGS}{31\xspace}
\newcommand{\tracemad}{0.055\xspace}      
\newcommand{\wavecalresel}{0.03--0.08\xspace}

\begin{document}

\title{
An Archival Calibration of JWST/MIRI Prism Wide-Field Slitless Spectroscopy: Methodology, Performance, and a Mid-Infrared Spectral Atlas of Galaxies at $z=0$--$4$ in the GOODS-N/S Fields
}

\author[0000-0002-4622-6617]{Fengwu Sun}
\affiliation{Department of Astronomy, School of Science, Westlake University, Hangzhou, Zhejiang 310030, P.\ R.\ China}
\affiliation{Center for Astrophysics $|$ Harvard \& Smithsonian, 60 Garden St., Cambridge, MA 02138, USA}
\email[show]{sunfengwu@westlake.edu.cn}

\author[0000-0002-7093-1877]{Javier \'{A}lvarez-M\'{a}rquez}
\affiliation{Centro de Astrobiología (CAB), CSIC-INTA, Ctra. de Ajalvir km 4, Torrejón de Ardoz, E-28850, Madrid, Spain}
\email[]{jalvarez@cab.inta-csic.es}

\author[0000-0003-1344-9475]{Eiichi Egami}
\affiliation{Steward Observatory, University of Arizona, 933 N Cherry Avenue, Tucson, AZ 85721, USA}
\email[]{egami@arizona.edu}

\author[0000-0002-7714-688X]{Rom\'{a}n Fern\'{a}ndez Aranda}
\affiliation{Centro de Astrobiología (CAB), CSIC-INTA, Ctra. de Ajalvir km 4, Torrejón de Ardoz, E-28850, Madrid, Spain}
\email[]{rfernandez@cab.inta-csic.es}

\author[0000-0003-4337-6211]{Jakob M. Helton}
\affiliation{Department of Astronomy and Astrophysics, The Pennsylvania State University, University Park, PA 16802, USA}
\email{jakobhelton@psu.edu}

\author[0000-0001-6052-4234]{Xiaojing Lin}
\affiliation{Department of Astronomy, Tsinghua University, Beijing 100084, People’s Republic of China}
\email[]{xiaojinglin.astro@gmail.com}

\author[0000-0002-6221-1829]{Jianwei Lyu}
\affiliation{Steward Observatory, University of Arizona, 933 N Cherry Avenue, Tucson, AZ 85721, USA}
\email[]{jianwei@arizona.edu}

\begin{abstract}
Through the Low Resolution Spectroscopy (LRS) observing mode of \jwst/MIRI, the P750L prism disperses the entire $\sim70''\times110''$ imager field, turning MIRI into a wide-field slitless spectrograph (WFSS) at 5--14\um\ with spectral resolution $R\sim100$.
Using archival MIRI LRS data taken in the GOODS-N/S deep fields and calibration programs, we present a complete and precise calibration suite for MIRI WFSS through an AI-assisted agentic workflow under human supervision.
The calibration includes flat field, sky background, spectral tracing, wavelength calibration, flux calibration, and the large-scale flux flat (L-flat).
We present the methodology of each calibration and quantify its accuracy.
As a demonstration, we extract 5--14\um\ slitless spectra of \Natlas\ spectroscopically confirmed galaxies at $z=0.08$--$3.71$ in the GOODS-N/S fields.
We detect a large number of polycyclic aromatic hydrocarbon (PAH) bands and atomic lines (including Paschen\,$\alpha$ out to $z=3.71$) in individual sources. 
The PAH 3.3\um\ luminosities correlate with the \paa\ luminosities (measured through NIRCam WFSS) for $z\sim1$ galaxies over a $\sim$1.5-dex range, supporting PAH 3.3\um\ as a star-formation-rate tracer out to $z\sim1.6$.
We release the calibration reference files, reduction scripts, and the mid-infrared spectral atlas.
We also present caveats and lessons learned from our archival calibration work.
Our empirical sensitivity characterization identifies a noise floor for deep MIRI WFSS integrations ($\gtrsim$ a few ks), whereas shallower integrations exceed the sensitivity predicted by the \jwst\ Exposure Time Calculator.
We also provide suggestions for the planning of future MIRI WFSS observing programs.
\end{abstract}


\begin{figure*}[!t]
\centering
\includegraphics[width=\linewidth]{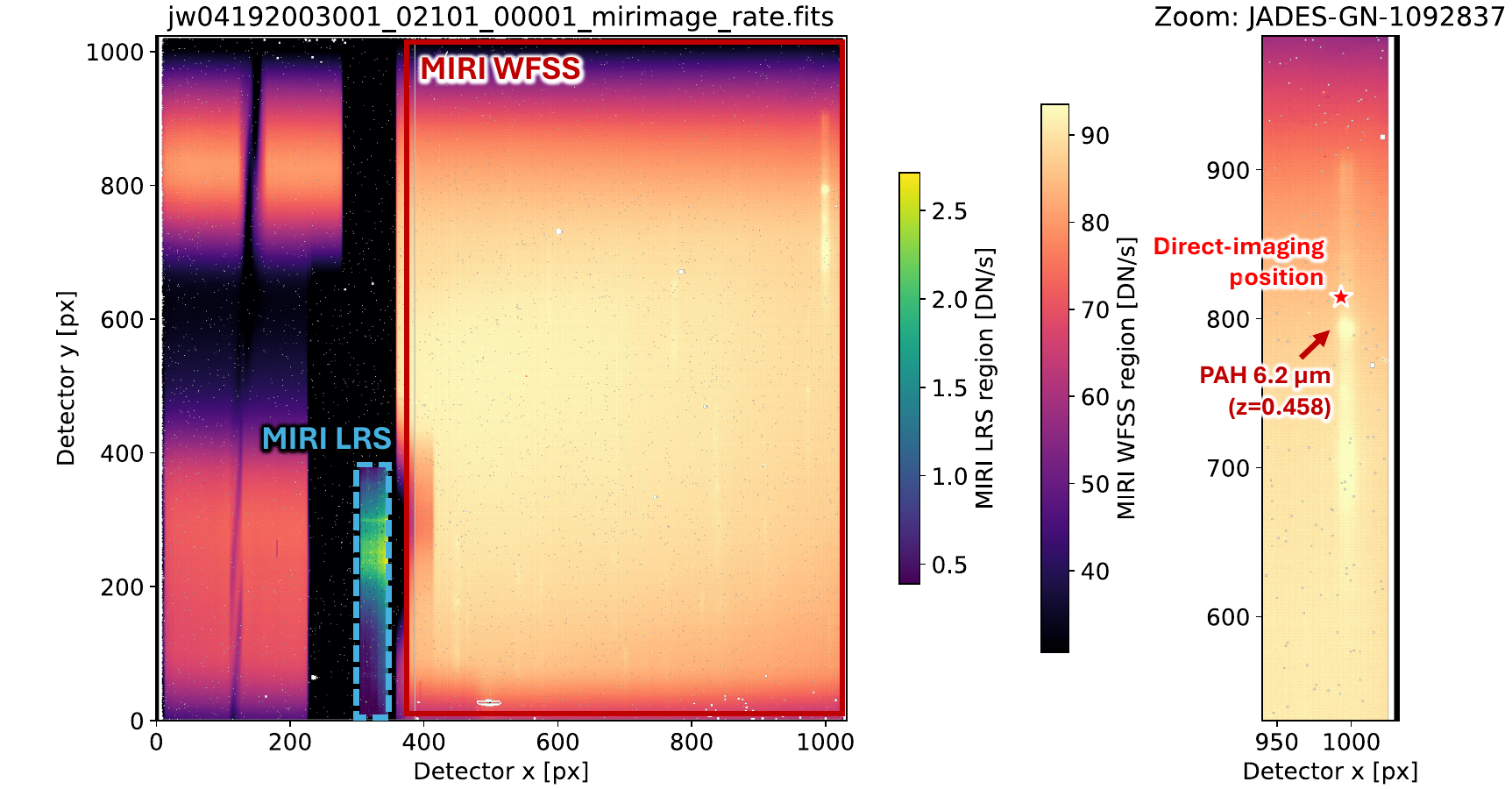}
\caption{
Overview of the JWST/MIRI WFSS configuration obtained through the P750L prism.
\textbf{Left}: the full-detector view of \texttt{jw04192003001\_02101\_00001}, a MIRI LRS exposure in the GOODS-N field with a 363.5\,s integration time, processed through the standard JWST stage-1 pipeline (i.e., a \texttt{\_rate} file). 
The image is displayed with two independent color stretches: the open imager field (``MIRI WFSS,'' red outline; FoV: $\sim$70\arcsec$\times$110\arcsec) and the coronagraphic field (left) are dominated by the bright dispersed thermal$+$zodiacal background, while outside them only the light entering the LRS slit reaches the detector (``MIRI LRS,'' dashed cyan), and the background is fainter by a factor of $\sim40$.
\textbf{Right}: zoom on the dispersed trace of the bright galaxy JADES-GN-1092837 ($z=0.458$). The red star marks its direct-image position, with wavelength increasing downward along the trace.
The redshifted PAH 6.2\,\micron\ feature is visible just below the red star (near $y\sim800$\,px).
\label{fig:mode}}
\end{figure*}

\section{Introduction}
\label{sec:intro}

The rest-frame mid-infrared spectra of galaxies carry unique and key information on their interstellar medium (ISM).
Broad emission bands from polycyclic aromatic hydrocarbons \citep[PAHs; e.g.,][]{leger1984, allamandola1985, tielens2008, li2020} at 3.3, 6.2, 7.7, 8.6, and 11.3\micron\ trace photodissociation regions and provide star-formation-rate (SFR) indicators that is highly robust against extinction \citep[e.g.,][]{calzetti2007, shipley2016}.
Atomic fine-structure lines, e.g., \arii\,6.99\micron, \neii\,12.81\micron, \siv\,10.51\micron, and \htwo\ lines probe the ionized and warm molecular gas, while hydrogen recombination lines (e.g., \paa, \bra) measure nearly unobscured star formation.
Prior to JWST, detailed mid-IR spectroscopic studies of galaxies started with the Infrared Space Observatory \citep[ISO; e.g.,][]{genzel1998, genzel2000}.
\spitzer/IRS then assembled this picture for luminous galaxies out to $z\sim2$--3 \citep[e.g.,][]{houck2004, armus2007, pope2008}, but its sensitivity confined detailed mid-IR spectroscopy to the brightest systems.

\jwst/MIRI \citep{rieke2015, wright2015, wright2023} re-opens this window with orders-of-magnitude improved sensitivity.
The Medium Resolution Spectroscopy \citep[MRS;][]{wells2015} mode delivers $R(\lambda)\sim1500$--3500 integral-field spectroscopy over a small ($\lesssim$$7''\times8''$) field, and Low Resolution Spectroscopy \citep[LRS;][]{kendrew2015} provides $R(\lambda)\sim40$--200 spectra at 5--14\,\micron\ through a $0\farcs51\times4\farcs7$ slit or in a slitless subarray mode, which was originally designed for time-series observations of bright sources.
However, neither the MRS nor the LRS mode is designed for blind spectroscopic surveys of faint galaxies over a wide field.

A third, largely unexploited configuration exists.
When the P750L prism is paired with the FULL-array MIRI readout, either intentionally or as the by-product of LRS slit observations, the prism effectively disperses the entire unobstructed imager field.
Under this circumstance, MIRI becomes a wide-field slitless spectrograph over the full $\sim70\arcsec\times110\arcsec$ imaging field of view (Figure~\ref{fig:mode}).
Every source in the field produces a $\sim$5--14\,\micron\ spectrum dispersed along the detector columns, in direct analogy to the NIRCam and NIRISS wide-field slitless spectroscopy (WFSS) modes that have transformed emission-line surveys at $\lambda<5$\micron\ \citep[e.g.,][]{greene2017, sunf2022, sunf2023, oesch2023}.
MIRI WFSS thus carries substantial scientific potential: a single MIRI pointing on a deep extragalactic field simultaneously yields mid-IR spectra of dozens of galaxies whose redshifts, stellar populations, and near-IR spectra are already known or taken from previous \hst\ and \jwst\ imaging-spectroscopic surveys.

Starting from JWST Cycle 5 (2026), MIRI WFSS is offered as an observing mode supported through STScI.
The full calibration of this mode is expected to be challenging: the trace curvature, dispersion solution, and prism throughput may all change with the field position, and the strong zodiacal plus thermal background \citep{rigby23b} in the MIRI imager must be removed without the benefit of on/off nodding.
Calibrating this mode therefore requires a dedicated, field-dependent solution for every component of the signal chain, and as of June 2026, such a full calibration suite of MIRI WFSS is not yet publicly available.
The recent calibration and pipeline development through STScI is presented by \citet{petric2026}, and a demonstration of this mode with observations in Hubble Ultra Deep Field (HUDF) is presented by \citet{kendrew2026}.

Fortunately, the JWST archive already contains the data needed for such a calibration.
As of June 1, 2026, \Nallexp\ FULL-array MIRI P750L exposures from \Nallprog\ programs are publicly available in the Mikulski Archive for Space Telescopes (MAST)\footnote{\url{https://mast.stsci.edu/}}.
Among them, several programs lie in cosmological deep fields such as the GOODS fields \citep{giavalisco2004}, where rich ancillary data exist for cross-validation, including JWST NIRCam imaging \citep[e.g., JADES;][]{rieke2023, eisenstein2023, johnson2026, robertson2026}, NIRCam grism spectroscopy (e.g., \fresco; \citealt{oesch2023}; CONGRESS; \citealt{Sun_CONGRESS}; and \sapphires; \citealt{sunf2025b}), MIRI imaging (e.g., SMILES; \citealt{alberts24a}; JADES, \citealt{alberts26a}; MEOW; \citealt{leung26a}) and NIRSpec multi-object spectroscopy \citep{deugenio2025_dr3, curtis-lake2026, scholtz2026}.
On the other hand, dedicated STScI calibration programs have observed a planetary nebula in the LMC as the wavelength calibrator (CAL-9505) and a CALSPEC \citep{bohlin2014} star \hdstar\ as a flux standard (CAL-9265).
Together these data can address the tracing, dispersion, and flux dimensions of the calibration problem.

This paper presents the complete calibration of MIRI P750L wide-field slitless spectroscopy (hereafter MIRI WFSS) built from these archival data, and demonstrates its capability with a spectral atlas of \Natlas\ galaxies at $z=0.1$--3.7 in the GOODS-N and GOODS-S fields.
Our calibration work flow combines the previous experience and lessons from the NIRCam WFSS calibration works \citep[][]{sunf2022, sunf2023, riekem2023, pirzkal2026}, as well as an AI-assisted agentic workflow, in which a large-language-model agent wrote, executed, iterated and optimized on the reduction and analysis code under careful human steering and review.
We describe the workflow, quantify the accuracy of every calibration component, and release the reference files, data processing pipeline, and the obtained spectral atlas of GOODS-N/S galaxies to the community.

The paper is organized as follows.
Section~\ref{sec:obs} describes the archival observations and ancillary data.
Section~\ref{sec:cal} presents the calibration methodology and the six calibration components with their validation.
Section~\ref{sec:atlas} applies the calibration to the GOODS fields and presents the mid-IR galaxy spectral atlas.
Section~\ref{sec:caveats} collects caveats and guidance for future MIRI WFSS observations.
A summary can be found in Section~\ref{sec:summary}.
All magnitudes are in the AB system \citep{oke1983}, and wavelengths are in vacuum.

\begin{figure*}[!t]
\centering
\includegraphics[trim={0 42pt 0 0},clip,width=\linewidth]{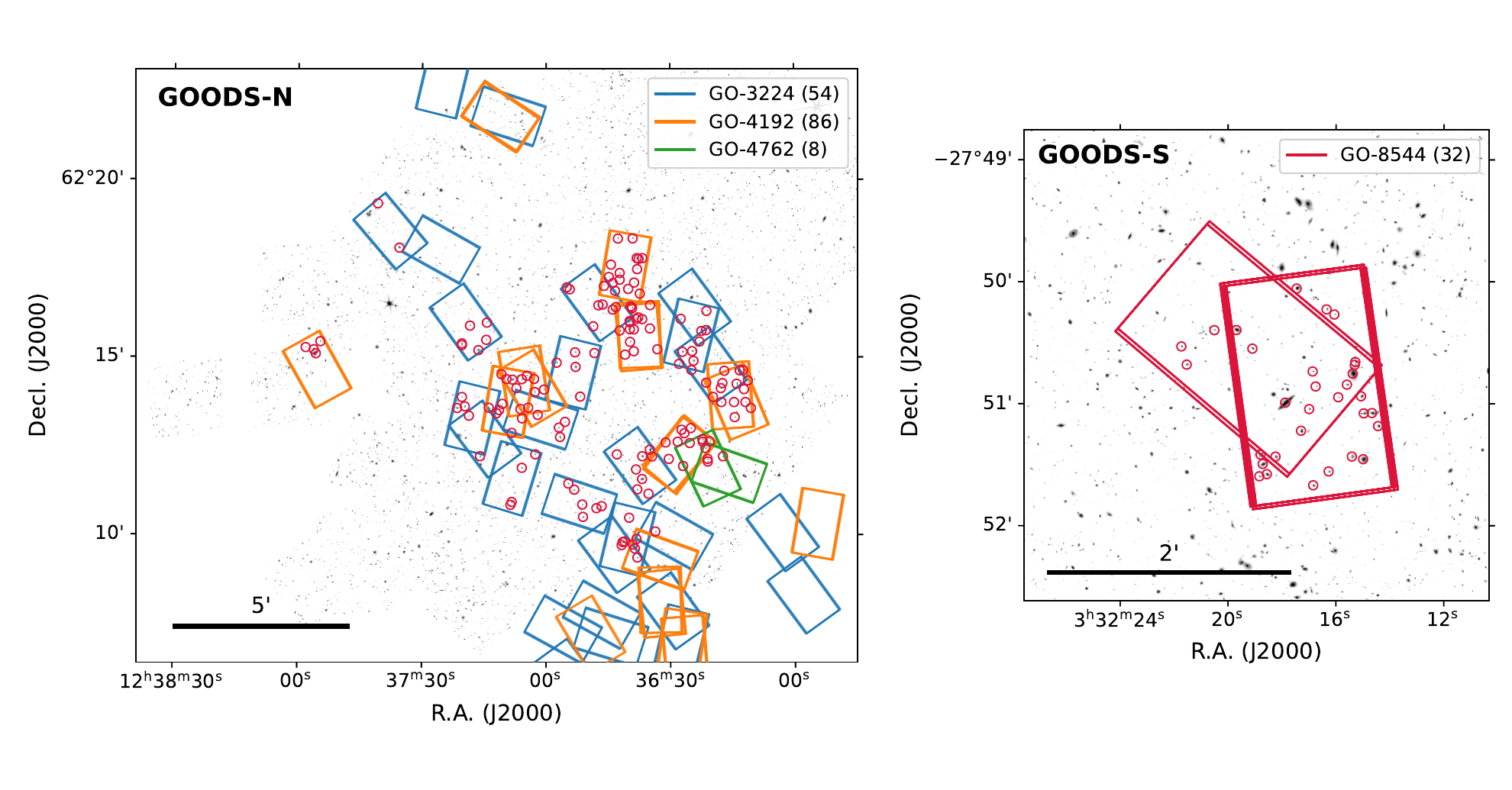}
\caption{
MIRI WFSS footprints over the JADES NIRCam F444W mosaics of GOODS-N (left) and GOODS-S (right).
Each rectangle is the illuminated WFSS region ($x=387$--1020, $y=15$--1017 in detector coordinates, $\simeq 70''\times110''$) of one exposure, color-coded by program.
Red circles mark the \Natlas\ galaxies of the spectral atlas (Section~\ref{sec:atlas}).
The scale bars are shown in the bottom-left corners.
\label{fig:footprints}}
\end{figure*}

\section{Observations and Data}
\label{sec:obs}


We queried the MAST archive for all public MIRI exposures taken with the P750L prism and the FULL imager readout. 
The census returns \Nallexp\ exposures from \Nallprog\ programs.
The majority are LRS slit observations of bright stars for exoplanet transit spectroscopy, a major science case for MIRI LRS \citep[][]{kendrew2015}.
In these observations, the dispersed fields are dominated by the background and serve here as the statistical basis for the flat field and master sky (Section~\ref{sec:cal_flat}--\ref{sec:cal_sky}).
A minority of the programs were analyzed more carefully as they carry key targets that can be used for tracing, wavelength and flux calibration, and we summarize these programs in Table~\ref{tab:programs}. 
We present an overview of these programs as follows.

\subsection{GOODS-N and GOODS-S pointings}
\label{sec:obs_goods}

We analyze archival MIRI WFSS observations taken in the GOODS-N/S fields \citep{giavalisco2004} as they observe a large number of stars and galaxies with known coordinates, mid-infrared flux densities and redshifts.
Figure~\ref{fig:footprints} shows the MIRI WFSS footprints of the four deep-field programs over the JADES NIRCam F444W mosaics \citep[made available through JADES DR5;][]{eisenstein2023, johnson2026, robertson2026}.
In GOODS-N, GO-3224 (PI: McKinney; \citealt{mckinney2026}) and GO-4192 (PI: Alberts) tile a wide area with 27 and 13 distinct pointings (typically 2--8 exposures each), while GO-4762 (PI: Fujimoto) concentrates 8 exposures on a single position near the quasar GNz7q \citep{fujimoto2022}.
In GOODS-S, GO-8544 (PI: Helton) places 32 exposures (68.1 hr) on a single pointing within the JADES Origins Field \citep{eisenstein2025} for galaxy JADES-GS-z14-0 \citep{carniani2024}, providing the deepest MIRI WFSS data on the sky taken so far \citep{helton2025}.

The wide-tiling strategy in GOODS-N yields typically 2--10 overlapping exposures per sky position at varying position angles, whereas the GOODS-S stack reaches 24--32 exposures at a nearly constant position angle.
The effective integration time per exposure ranges from 111 to 1812\,s (median 433\,s) across the GOODS-N field, whereas all 32 GOODS-S exposures share a uniform 7656\,s ($2.13$\,hr) integration.

\subsection{LMC wavelength-calibration field (CAL-9505)}
\label{sec:obs_lmc}

CAL-9505 (PI: Petric; \citealt{petric2026}) obtained the dedicated \texttt{MIR\_WFSS} calibration observations on a crowded field in the Large Magellanic Cloud (LMC) centered on the planetary nebula \lmcpn.
The four dithered exposures disperse hundreds of stars with well-determined Gaia DR3 astrometry \citep{gaiadr3}, and the planetary nebula provides a series of narrow emission lines. 
These data are analyzed for anchoring the spectral tracing (Section~\ref{sec:cal_trace}) and the wavelength solution (Section~\ref{sec:cal_wave}).

\subsection{CALSPEC standard star grid (CAL-9265)}
\label{sec:obs_hd}

CAL-9265 (PI: Petric; \citealt{petric2026}) observed \hdstar, an A1V CALSPEC standard \citep{bohlin2014, bohlin2020} with Gaia $G=6.83$.
The observations were conducted in a five-position grid across the WFSS field, with 11 intra-pixel dither scans per position (55 exposures).
In these observations, the standard star lands within the unobstructed WFSS region in all 55 exposures, providing a direct, model-independent measurement of the flux response at five widely separated field positions.
However, the core of the stellar trace saturates at $<$7.4\um, and we only use the 7.4--13.5\um\ component for flux calibration (see Section~\ref{sec:cal_flux}).

\subsection{Data processing}
\label{sec:obs_processing}

We start from the standard \texttt{jwst} pipeline stage-1 rate images (\texttt{rate.fits}) retrieved from MAST.
Each rate image is divided by the flat field (Section~\ref{sec:cal_flat}) and cleaned of the zodiacal plus thermal sky background (Section~\ref{sec:cal_sky}).
World coordinate solutions (WCS) are attached with the standard \texttt{assign\_wcs} step assuming the MIRI imaging instrumental aperture, which is accurate for the undispersed coordinate and based on the assumption that the astrometry has been tied to Gaia.
Such an assumption should remain valid if the archival data are obtained through MIRI LRS observations after proper and successful target acquisition (however, see caveats in Section~\ref{sec:caveats_astrometry}).
We refer to these intermediate products as level-1.5 images, similar to the products defined by \citet{sunf2023} for NIRCam WFSS images after similar processing steps.
All spectral extraction in this paper is conducted directly on the level-1.5 images in units of \dnps.

\subsection{Ancillary data}
\label{sec:obs_ancillary}

A large number of deep ancillary data exist in the GOODS fields, and we use them for source selection, cross-dispersion profile modeling and validation.
Photometry and astrometry come from the JADES DR5 NIRCam catalogs and mosaics \citep[][]{eisenstein2023, johnson2026, robertson2026}.
Spectroscopic redshifts are drawn from curated field compilations (\citealt{puskas2025a}).
The redshifts include those from literature ground-based observations, MUSE spectroscopy (primarily from \citealt{urrutia2019}), JWST/NIRSpec primarily from JADES DR4 \citep{curtis-lake2026,scholtz2026}, as well as NIRCam grism WFSS \citep{oesch2023, linx2026a, sunf26}.
NIRCam grism spectra at 3.1--5.0\um\ from \congress\ (F356W, GOODS-N; \citealt{Sun_CONGRESS}), \fresco\ (F444W, GOODS-N/S; \citealt{oesch2023}), and \sapphires\ (F444W, GOODS-N/S; \citealt{sunf2025b}) provide an independent flux scale that connects the blue end of the MIRI WFSS coverage (Section~\ref{sec:atlas_sed}).

\begin{figure*}[!t]
\centering
\includegraphics[width=\linewidth]{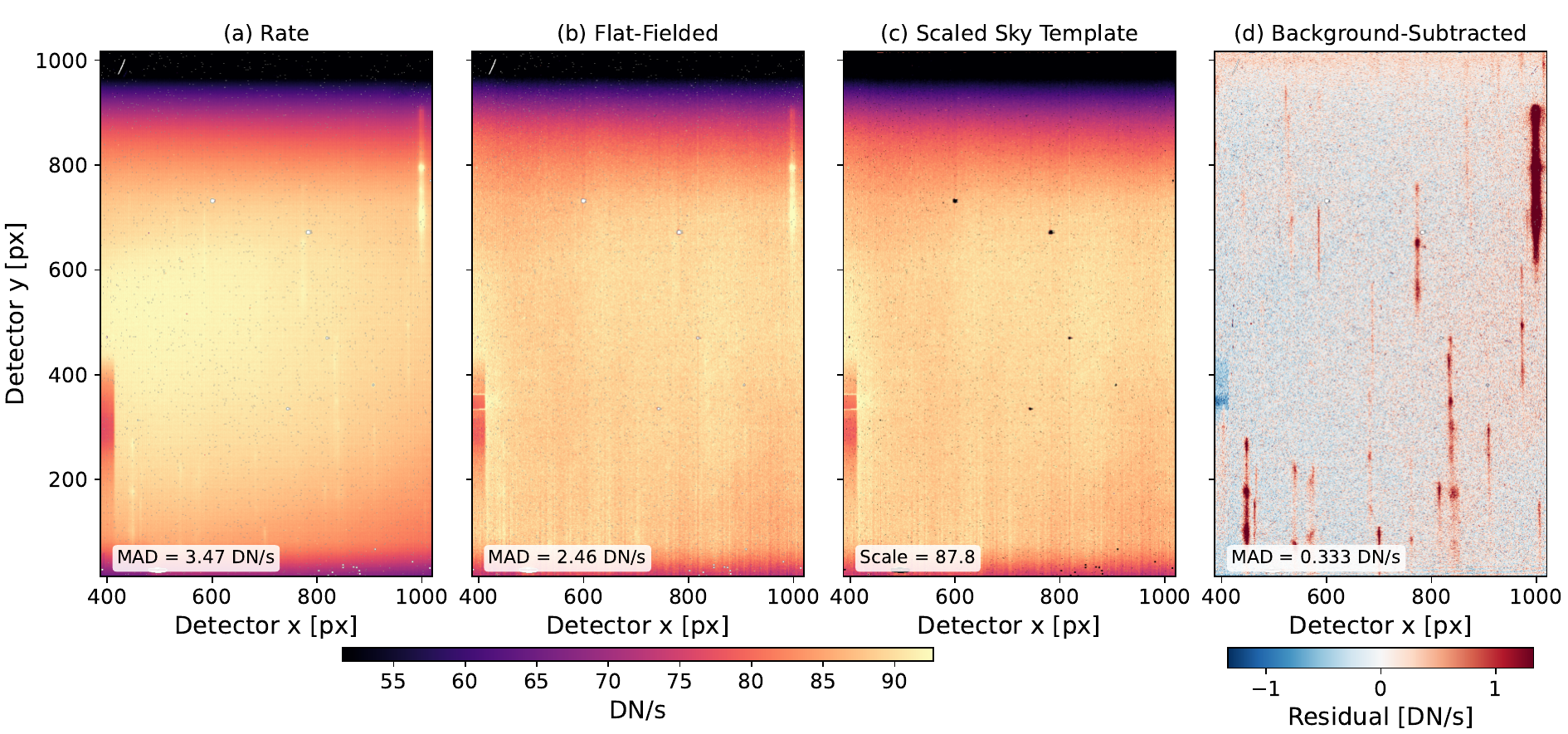}
\caption{
The image-calibration sequence on the MIRI WFSS region of a representative GOODS-N exposure (the same as in Figure~\ref{fig:mode}).
From left to right: (a) the rate image; (b) after flat-fielding and subtraction of the additive defect map; (c) the matched background model (master-sky template scaled to the frame); (d) the background-subtracted level-1.5 frame, including the per-row clipped-median correction.
Tens of dispersed spectra of bright sources are clearly visible in panel (d).
Panels a--c share the color stretch and bottom-left color bar. 
The in-panel annotations give the robust scatter (MAD) of the source-masked WFSS region in each step as well as the fitted template scale.
\label{fig:lv15seq}}
\end{figure*}

\section{MIRI WFSS Calibration}
\label{sec:cal}

The observed count rate of a source at direct-image position $(x_0, y_0)$ follows the model
\begin{equation}
\label{eq:model}
\begin{split}
{\rm DN\,s^{-1}}(x_s, y_s) = P(x_s, y_s) \big[ &f_R(\lambda)\, L(x_0, y_0)\, F_\nu(\lambda) \\
&+ {\rm sky}(x_s, y_s) \big],
\end{split}
\end{equation}
where $P(x_s, y_s)$ is the pixel flat field (Section~\ref{sec:cal_flat}), which multiplies both the source signal and the dispersed sky background ${\rm sky}(x_s, y_s)$ (Section~\ref{sec:cal_sky}, in flat-fielded units), $f_R(\lambda)$ is the spectral response (Section~\ref{sec:cal_flux}), and $L(x_0, y_0)$ is the large-scale flux flat (L-flat; Section~\ref{sec:cal_lflat}).
The dispersed spectral coordinates $(x_s, y_s)$ map to wavelength $\lambda$ through the field-dependent trace and dispersion solutions (Sections~\ref{sec:cal_trace}--\ref{sec:cal_wave}).
Throughout this section, the dispersion runs along the detector $y$ axis (longer wavelengths toward more negative $\Delta y = y_s - y_0$), and the illuminated WFSS region spans $x=387$--1020, $y=15$--1017 with a pixel size of 0\farcs11/pixel.
This section describes the derivation and validation of each component.

\subsection{AI-assisted calibration workflow}
\label{sec:cal_method}

The calibration presented here was developed in a workflow assisted by an artificial intelligence (AI) agent.
A large-language-model coding agent (Claude Opus 4.8 \& Fable 5; Anthropic) wrote, executed, debugged, and iterated on all reduction and analysis software.
The agent navigated the MAST archive, chose and adjusted algorithmic strategies, and produced the diagnostics under continuous direction and inspection by the author.

We derive the calibration components in the order in which they depend on one another, verifying each before the next is built upon it.
We first establish the image-level corrections.
These include the flat field (Section~\ref{sec:cal_flat}) and the sky background (Section~\ref{sec:cal_sky}), which are used to reduce every stage-1 rate image to level-1.5 (Section~\ref{sec:obs_processing}).
We then model the spectral tracing (Section~\ref{sec:cal_trace}) and the dispersion (i.e., wavelength calibration; Section~\ref{sec:cal_wave}).
These geometric calibrations are drawn heavily from human input and based on the prior experience with NIRCam WFSS commissioning and calibration \citep{sunf2022, sunf2023, riekem2023, pirzkal2026}, for example, selecting proper spectral traces and measuring their centroids.
With the geometry fixed, we measure the spectral response $f_R(\lambda)$ (i.e., flux calibration; Section~\ref{sec:cal_flux}) and the field-dependent large-scale flat field (i.e., L-flat; Section~\ref{sec:cal_lflat}).
These calibrations were obtained based on the CALSPEC standard-star grid and bright stars with known mid-IR flux densities in the GOODS-N field.

Wherever possible, we verify each calibration component against measurements that do not enter its derivation, so a defective solution is detected before it propagates.
These validations are presented in the corresponding subsections below and in Section~\ref{sec:atlas}.
This human-in-the-loop review caught a number of genuine errors from the AI agent.
For example, we identified an error in the intermediate flux calibration because the AI agent did not recognize the saturated spectral pixels from the bright flux calibrator.
Through our human reviews, such errors were fixed over multiple iterations of steering.
Section~\ref{sec:cal_release} presents the calibration references released with this work.

\subsection{Flat field}
\label{sec:cal_flat}

The P750L prism transmits the full 5--14\um\ band onto every MIRI pixel.
Therefore, the appropriate flat field is a broadband average weighted by the local spectral energy distribution (SED) of the background.
Because the MIRI pixel response is nearly achromatic, we tested the CRDS imaging flats of the nearby filters (F560W and F770W), retrieved under the CRDS context \texttt{jwst\_1337.pmap}.
We also built a ``native'' flat directly from the dispersed data.
This flat field is a sigma-clipped stack of the background-dominated P750L frames.
Even though we have masked bright spectral traces prior to stacking, this native flat is contaminated, as the dispersed-source streaks survive the stacking and imprint low-level vertical striping.
Therefore, we discard the native flat because of a large RMS noise.
We then divide the background-dominated frames by each candidate flat and compare the resulting sky RMS.
The F560W flat yields the least residual structure (Figure~\ref{fig:lv15seq}, panel b), and therefore we adopt it.
We re-normalize it to unit median over the WFSS region and release it as the WFSS flat field (\texttt{flat\_P750L\_F560W.fits}).

\subsection{Sky background}
\label{sec:cal_sky}

The dispersed sky background of the MIRI imager is bright ($\sim 10^2$ \dnps\ per pixel), spatially structured, and relatively stable in pattern but variable in amplitude.
We construct a master sky background model by sigma-clipped stacking of the flat-fielded archival FULL P750L frames, after masking detected traces and scaling each frame to a common background level.

The sky background subtraction proceeds in two modes.
Mode A scales the master sky to each science frame (one multiplicative degree of freedom) and subtracts it.
This is our default mode, and it is fully linear.
Figure~\ref{fig:lv15seq} shows the full sequence of flat-fielding and sky background subtraction on a representative exposure.
Mode B additionally fits a five-component principal component analysis (PCA) basis of the residual sky variations, simultaneously with the template amplitude, by linear least squares.
The basis is derived from the same archival stack.
Imperfectly masked sources leave compact imprints in the components, and we interpolate over these regions so the basis captures only sky variations.
We compare these two modes in Section~\ref{sec:caveats_mode}.
After mode-A subtraction, the source-masked residual background is 1.25--1.33$\times$ the propagated pixel noise at the $\sim$450\,s tiling depth, and higher in longer integrations.
This excess is a fixed-pattern calibration systematic that does not average down, so we treat it as a noise floor rather than rescaling the per-pixel uncertainties (see further discussion in Section~\ref{sec:caveats_depth}).

The top of the WFSS region ($y \gtrsim 940$) requires special care.
The dispersed background there is faintest (0.4--0.6$\times$ the central level) because sources and sky outside the pick-off mirror cannot produce the red part of the spectral traces there.
To properly mask source traces in this region, we use a median filter to remove the background prior to trace detection.

We also refine the sky background template against the data themselves.
To do so, we take the median of the source-masked residuals across all 304 calibrated science frames from different fields.
This isolates the systematic errors that persist at fixed detector positions, e.g., warm pixels that the stacking had excluded.
Together these pixels affect 1.5\% of the total MIRI WFSS FoV, at amplitudes of up to tens of \dnps.
A per-pixel regression of these residuals against the per-frame sky scale separates a multiplicative part from an additive part.
The multiplicative part is a genuine template error, and we fold it back into the template.
The additive part consists of detector defects at constant \dnps, independent of the sky level.
We release it as a separate defect map, to be subtracted unscaled, and flag the strongest defects ($>$0.5 \dnps) as unusable.

After subtracting the template, we compute a sigma-clipped median of the residual for every detector row using the source-masked background pixels of the WFSS region, and remove it.
This per-row correction (median absolute amplitude $\sim$0.06 \dnps) suppresses the electromagnetic-interference (EMI) banding of the FASTR1 readout.
As a known MIRI artifact \citep{brandt25_miriemi}, the EMI banding is not phase-coherent between exposures, so it averages out of the background template but not out of individual frames.
The correction also removes any residual row-wise gradient that a single scaled template cannot follow, and it improves the top-region residual median absolute deviation
(MAD) by a further $\sim$10\%.

\begin{figure*}[!t]
\centering
\includegraphics[width=\linewidth]{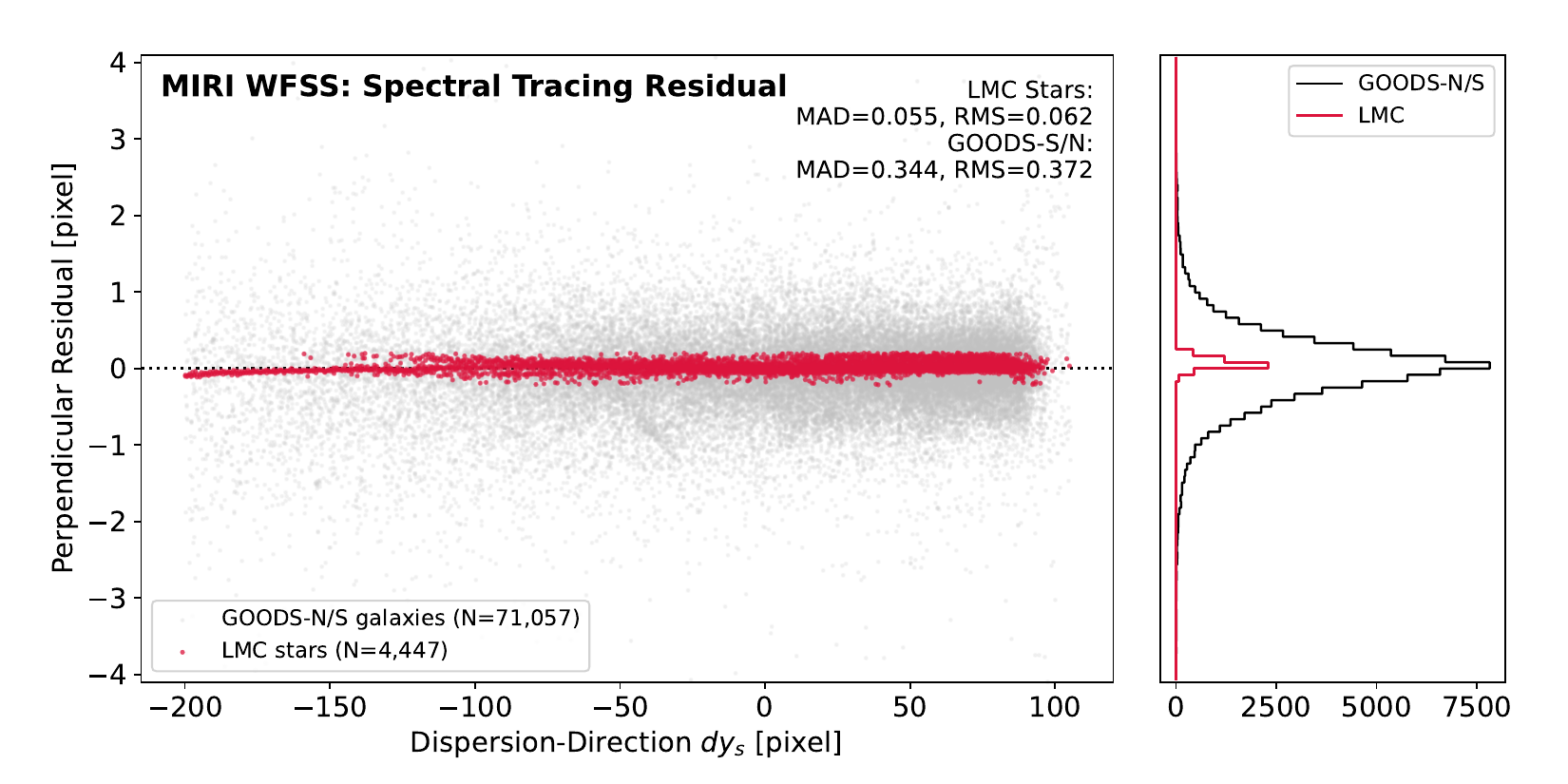}
\includegraphics[width=\linewidth]{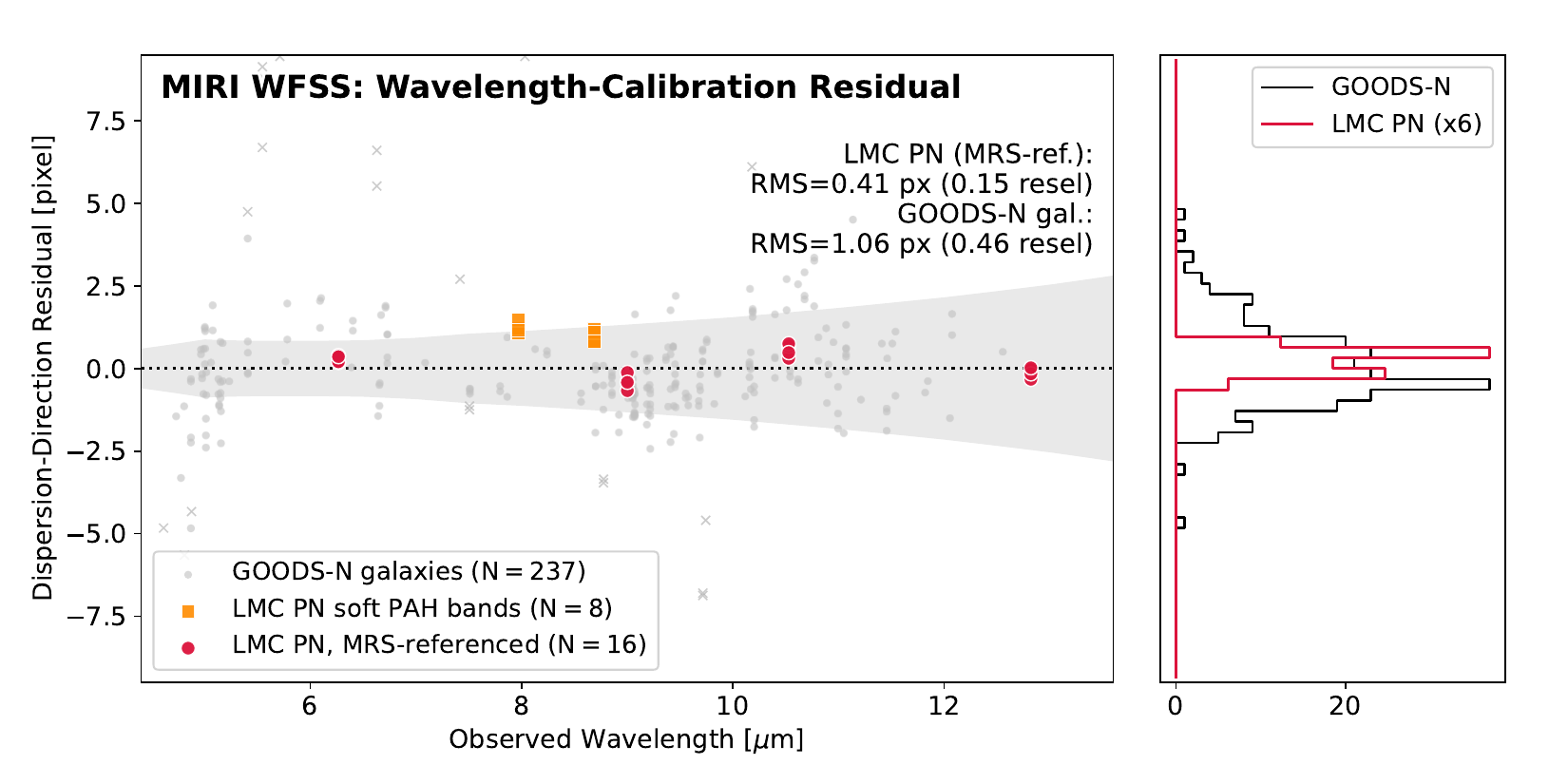}
\caption{
The spectral tracing (top) and wavelength solution (bottom) calibration residuals against the anchors, both measured in detector pixels.
\textbf{Top:} cross-dispersion residual of the global trace model versus dispersion offset ($dy_s$). Measurements include the proper-motion-corrected LMC stars (red; the primary anchors) and the GOODS field galaxy sample (gray).
The residual of our best-fit tracing model reaches a median absolute deviation of \tracemad\ pixel (RMS 0.062 pixel) against the LMC stars.
\textbf{Bottom:} dispersion-direction residual of the wavelength solution versus observed wavelength, for the \lmcpn\ narrow lines (red; RMS\,=\,0.12 pixel, 0.04 resolution element) and PAH bands (orange).
Residuals of the spectral features of GOODS-N galaxies (gray; RMS\,=\,0.97 pixel) and the $\pm$0.5-resolution-element band (gray shade) are also shown.
The right-hand panels give the marginal residual distributions. 
Faint crosses in the lower panel denote wavelength anchors that are rejected by the iterative sigma clipping.
\label{fig:trace}}
\end{figure*}

\subsection{Spectral trace}
\label{sec:cal_trace}

The spectral tracing model gives the cross-dispersion offset $\Delta x = x_s - x_0$ of the spectrum $(x_s, y_s)$ relative to the source position $(x_0, y_0)$ as a function of field position and dispersion offset:
\begin{equation}
\Delta x(x_0, y_0, \Delta y) = A(x_0, y_0) + B(x_0, y_0)\,\Delta y + C(x_0, y_0)\,\Delta y^2,
\end{equation}
with $A$, $B$, $C$ each a two-dimensional cubic polynomial in the source position (30 coefficients in total), and $\Delta y = y_s - y_0 \in [-262, +150]$ pixels.
We fit the model to centroid measurements of the traces of bright sources across the GOODS-N, GOODS-S, and LMC fields.
Gaia DR3 \citep{gaiadr3} stars in these fields have been corrected for proper motions.
We weight each measurement by the inverse variance of its centroid uncertainty. 
This naturally favors the LMC stars, as their median centroid uncertainty is 0.015 pixel, versus 0.12 pixel for the mostly extended galaxies in the GOODS fields. 
Throughout the fitting, we also conducted iterative rejection of contaminated rows that show large residuals in $\Delta x$.
The resultant global fit achieves a MAD of \tracemad\ pixel (RMS 0.062 pixel) against the LMC stars (Figure~\ref{fig:trace}, top panel).


\begin{figure*}[!th]
\centering
\includegraphics[width=\linewidth]{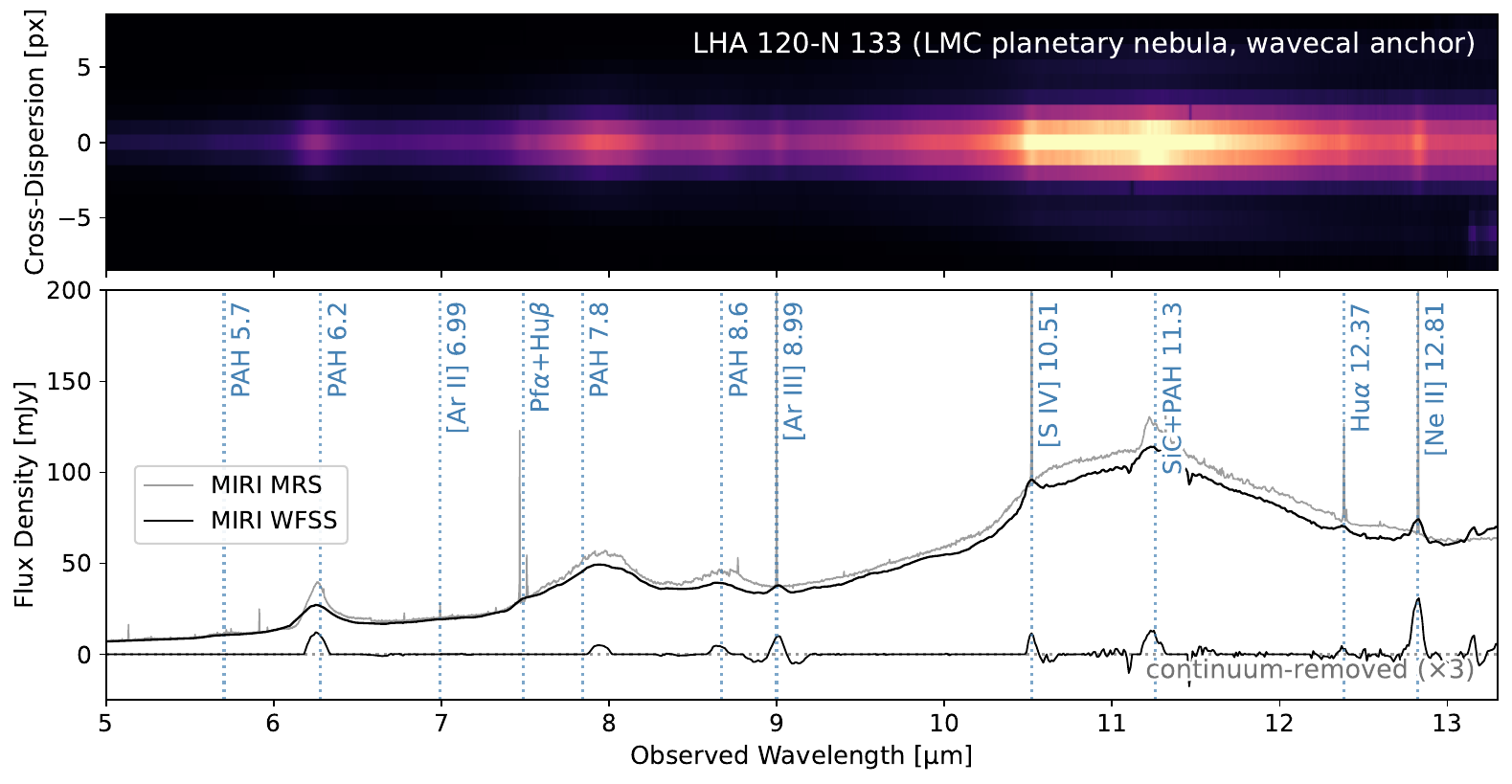}
\caption{
The fully calibrated MIRI WFSS spectrum of \lmcpn, co-added over the four dithered exposures of CAL-9505.
The 2D spectrum is shown in the top panel and the 1D spectrum in the bottom.
The identified spectral line features are labeled with their vacuum wavelengths.
The gray curve shows the continuum-removed 1D spectrum (scaled by $\times$3 for clarity), in which the narrow anchor lines used for the wavelength calibration stand out (Section~\ref{sec:cal_wave}).
\label{fig:pn_calibrated}}
\end{figure*}

\subsection{Wavelength calibration}
\label{sec:cal_wave}

The dispersion model gives the dispersion offset of each wavelength,
\begin{equation}
\Delta y(x_0, y_0, \lambda) = \sum_{i\le3} D_i(x_0, y_0)\, (\lambda - 3.95\,\micron)^i,
\end{equation}
Each coefficient $D_i(x_0, y_0)$ is a polynomial surface of the direct-imaging position, with decreasing polynomial order (16 coefficients in total).
The 3.95\um\ pivot is a convention carried over from our NIRCam WFSS tooling, as the NIRCam grisms have their undeflected wavelength at 3.95\,\micron\ \citep{riekem2023}. 
This choice does not affect the fitted solution or its accuracy.
We adopt the NIRCam pivot so that the same evaluation code serves both instruments.
The undeflected wavelength ($\Delta y = 0$) of the P750L WFSS configuration is 8.24\um\ at the center of the WFSS region, increasing to $\simeq$8.6\um\ toward the field corners.

The wavelength anchors come from both calibration observations and archival science observations.
The planetary nebula \lmcpn\ (or SMP LMC 058; CAL-9505) provides 24 anchor measurements at four dither positions, referenced to the MIRI/MRS spectrum of the same nebula \citep{jones2023} degraded to the LRS line spread function \citep{kendrew2015}.
Within each dither and feature-centered window, we cross-correlate the WFSS spectrum against the degraded MRS spectral template.
Both spectra were continuum-removed through identical median filtering before the cross-correlation, so the wavelength anchors are in sensitive to either the continuum shape or the aperture differences in spectral extraction.
This yields 16 solid wavelength anchors from four narrow features (PAH 6.2\um, \ariii\,8.99, \siv\,10.51, and \neii\,12.81\um), plus eight soft anchors from the PAH 7.8 and 8.6\,\micron\ band heads with deliberately inflated uncertainties (Figure~\ref{fig:pn_calibrated}).

We also include wavelength anchors from bright galaxies with known spectroscopic redshifts in the GOODS-N archival observations.
All galaxies and their spectral features are identified through visual inspection, and they contribute 272 additional line measurements through the \paa\ line and the PAH 3.3, 6.2, 7.7, 8.6 and 11.3\,\micron\ features.
These archival anchors from galaxies fill the field coverage.

Each wavelength anchor is weighted by the inverse variance of its total uncertainty.
The PN narrow-feature anchors carry their per-dither empirical uncertainties with a 0.30 pixel floor, and they are exempt from the iterative sigma clipping of the fit. 
The PN PAH band-head anchors carry an additional 1.0\,pixel systematic uncertainty added in quadrature, acting as soft constraints, because their centroid wavelengths are less accurate than those of the narrow features.
For galaxy anchors, we add per-wavelength-bin systematic floors in quadrature (0.3--3\,pixel), self-calibrated from their post-fit scatter, which account for their extended sizes and broad spectral features and automatically demote them where they are unreliable.
A PN narrow-feature anchor thus carries $\sim$10--100 times the weight of a galaxy anchor, depending on the wavelength bin.

The wavelength solution is accurate to 87--144 km s$^{-1}$ RMS over 7.9--12.8\um, i.e.\ \wavecalresel\ of a spectral resolution element.
It degrades to $\simeq$0.2 of a resolution element at the blue end (5.6--6.2\um), where the prism resolving power is lowest and anchor lines are scarce (Figure~\ref{fig:trace}, lower panel).
As an independent check, we recover the PAH 11.3\um\ complex of the PN at a flux-weighted centroid of 11.27--11.28\um\ in all four dithers.
Given the uncertainty in the PAH wavelength centroid, we do not include it for the fit while label it out as a validation feature. 
Further validation of the wavelength accuracy is presented in Section~\ref{sec:atlas}, where we stack the PAH and \paa\ spectra of galaxies with known spectroscopic redshifts.

\begin{figure*}[!t]
\centering
\includegraphics[width=\linewidth]{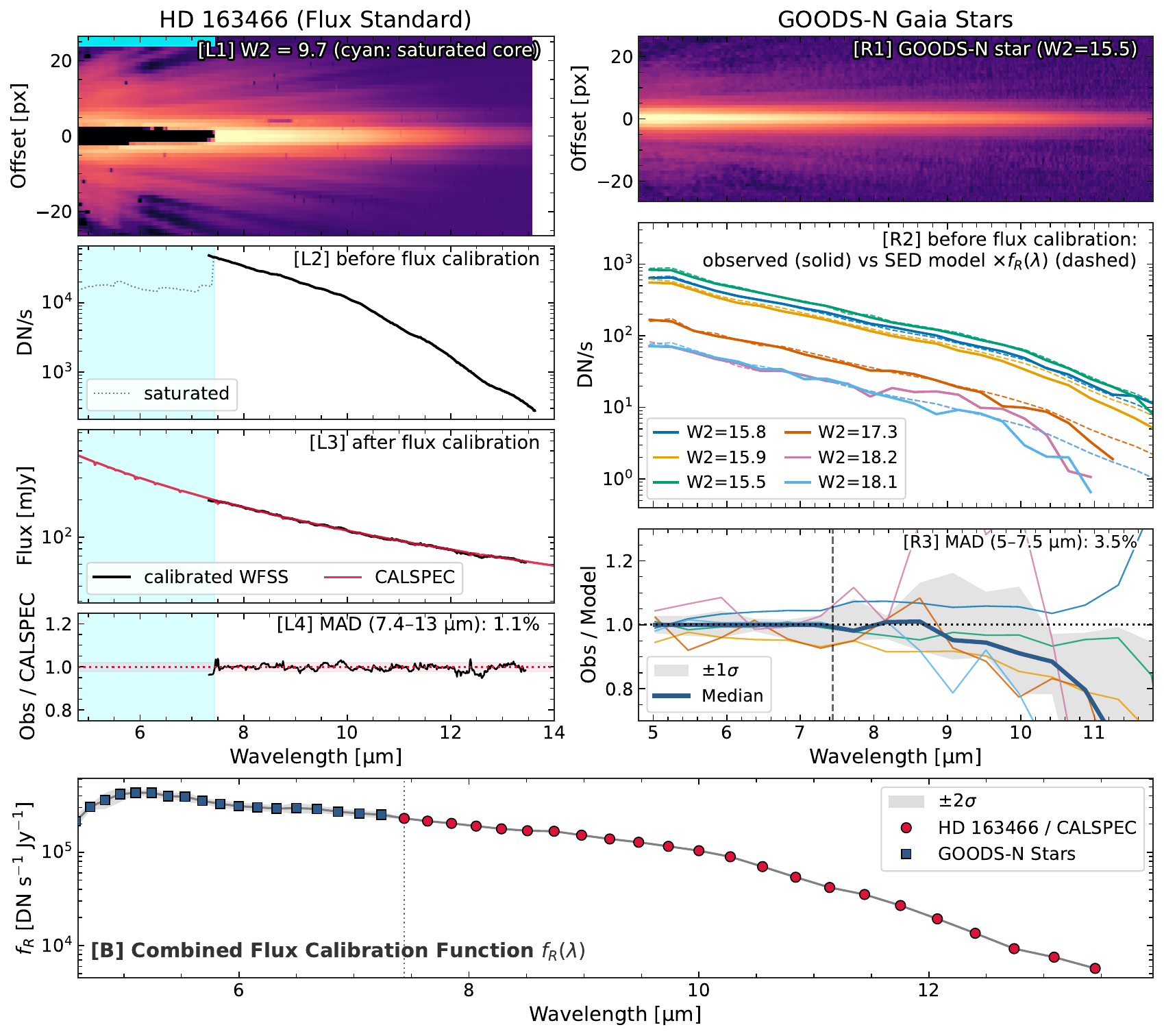}
\caption{
The flux calibration of JWST/MIRI WFSS.
\textbf{Left}: the flux calibration with the CALSPEC standard \hdstar\ (an A1V star), demonstrated through one representative exposure.
Panel L1: the rectified 2D spectrum (cyan marks the saturated core).
L2: the extracted count-rate 1D spectrum with and without saturation masking.
L3: the flux-calibrated spectrum against the CALSPEC model.
L4: the ratio of the calibrated spectrum to the CALSPEC model (MAD 1.1\% over 7.4--13\um).
\textbf{Right}: the flux calibration with six GOODS-N field stars, using independent SED modeling.
Panel R1: the rectified 2D spectrum of one bright ($G=15.6$) GOODS-N star before flux calibration, in the same style as panel L1.
R2: the per-star count-rate spectra, observed (solid) versus each star's modeled SED multiplied by the spectral response curve (SED\,$\times f_R$, dashed), spanning a $\sim$10$\times$ range in stellar brightness.
R3: the resulting per-star observed/modeled flux ratios (thin solid lines) with the ensemble median (thick blue line) and the star-to-star 16--84\% scatter (gray shading).
\textbf{Bottom} (panel B): the final flux-calibration function $f_R(\lambda)$, jointly constrained by the two columns of panels above.
The red circles are the measurements based on the CALSPEC \hdstar\ spectra (7.44--13.45\um), and the blue squares are based on the GOODS-N ensemble shape over 4.7--7.44\um.
The flux-calibration function is shaded with the $\pm2\sigma$ uncertainty.
\label{fig:fluxcal}}
\end{figure*}

\subsection{Flux calibration}
\label{sec:cal_flux}

We define the response $f_R(\lambda)$ as the aperture-summed count rate per unit flux density.
$f_R(\lambda)$ is therefore the count rate (\dnps) summed over a $\pm$20 pixel cross-dispersion aperture per dispersion row, divided by $F_\nu$ in Jy.
$f_R(\lambda)$ is measured through the spectra of the STScI CALSPEC standard star and bright GOODS-N stars, accounting for the aperture loss.

\subsubsection{CALSPEC standard}

We reduce the five-position grid of \hdstar\ spectra (CAL-9265), and divide each extracted 1D spectrum by the CALSPEC model spectrum \citep{bohlin2020} 
to obtain a per-exposure response measurement.
However, the spectral trace of \hdstar\ is saturated at $\lambda \lesssim 7.4$\um, and therefore we exclude this wavelength regime from the final calibration (Figure~\ref{fig:fluxcal}, panels L1--L4).
The 55 exposures constrain the per-wavelength-bin response to a formal precision of $\sigma(f_R)/f_R \simeq 0.3\%$ (standard error of the median) over 7.4--13.5\um.
The larger 1.1\% scatter of a single calibrated visit against CALSPEC (Figure~\ref{fig:fluxcal}, panel L4) reflects per-visit noise and residual systematics rather than the precision of the mean response.
The absolute accuracy is limited by the CALSPEC system itself \citep[1--2\%;][]{bohlin2014, gordon2022}.

\subsubsection{GOODS-N stars}

Blueward of the CAL-9265 saturation limit, we measure the response $f_R(\lambda)$ from an ensemble of bright Gaia stars in the GOODS-N field (Figure~\ref{fig:fluxcal}, right).
We model their intrinsic spectra by fitting PHOENIX stellar spectral models \citep{husser2013} to their 0.4--8\um\ SEDs, including HST/ACS, JWST/NIRCam, \spitzer/IRAC, 2MASS, and WISE.
Saturated JWST photometry is excluded from the fitting.
We exclude cool M dwarfs, because their mid-IR spectral extrapolations are likely unreliable.
Only well-fit, warm stars ($\chi^2_\nu < 20$, $T_{\rm eff} \ge 3400$ K) enter the ensemble, and the GOODS-N stars used for this calibration are displayed in Figure~\ref{fig:fluxcal} (panels R1--R3).

The flux calibration from GOODS-N stars is derived in the same manner as that with the CALSPEC standard.
We note that we rescale the ensemble shape to match the CALSPEC baseline in the 7.5--9.0\um\ overlap by a small factor of 1.053. 
The GOODS-N star ensemble fills the 4.7--7.4\um\ regime with $\sigma(f_R)/f_R \simeq 5\%$ (Figure~\ref{fig:fluxcal}, panel R3). 

\subsubsection{Aperture loss}

We also quantify the aperture loss of the $\pm$20 pixel window.
Collapsed along the dispersion axis, the MIRI point-spread-function models from \texttt{STPSF} \citep{perrin2014} place 90--94\% of the total light inside this window (0.899, 0.913, and 0.940 at F560W, F770W, and F1000W, relative to a 40\arcsec\ aperture).
This fraction is therefore absorbed bin-by-bin into $f_R(\lambda)$ for point sources. 
We validate the model profile directly on the dispersed standard at 9.5--10.5\um, where it is unsaturated.
Across the 55 grid exposures, the measured cross-dispersion curve of growth matches the \texttt{STPSF} prediction to within 0.9\% at every radius from $\pm$5 to $\pm$30 pixel (normalized at $\pm$40 pixel).
The empirical LRS point-spread functions released through CRDS (\texttt{jwst\_miri\_psf\_0002} and \texttt{0015}; \citealt{petric2025}) corroborate this agreement.


The bottom panel of Figure~\ref{fig:fluxcal} shows the final spectral response curve $f_R(\lambda)$, jointly constrained by the CALSPEC standard and the GOODS-N star ensemble.
$f_R$ peaks at $\sim 4.4\times10^5$\,\dnps\,Jy$^{-1}$ at $\lambda = 5.1$\um\ and falls by a factor of $\sim$80 to 13.45\um.
Our released calibration covers 4.7--13.45\um, and we advise against extrapolating the response curve.
This is because blueward of 4.7\,\micron\ the zeroth-order image overlaps with the spectrum, and no wavelength calibrator constraint exists redward of 13.45\,\micron.

\subsubsection{Validation}
\label{sec:cal_validation}

We conduct an end-to-end test of the flux calibration with \lmcpn, which is not used in any flux-calibration fit.
We obtain the fully calibrated co-added spectrum of \lmcpn\ as shown in Figure~\ref{fig:pn_calibrated}.
This spectrum recovers the expected continuum shape of the PN.
We then obtain the synthetic F560W photometry of this spectrum, and compare it with the aperture photometry measured on the F560W direct verification image of the same CAL-9505 visit.
The spectrum-to-image flux ratios (1.17 and 1.14 in $2\farcs2$ and $3\farcs0$ apertures) match the \texttt{STPSF}-predicted aperture corrections (1.14 and 1.11).
The aperture-corrected closure is thus 1.03 in both apertures, consistent with unity at the $\sim$3\% level given the CALSPEC-tied absolute scale and the photometric uncertainties, validating our flux calibration and aperture-loss treatment.


\subsection{L-flat}
\label{sec:cal_lflat}

\begin{figure}
\centering
\includegraphics[width=\linewidth]{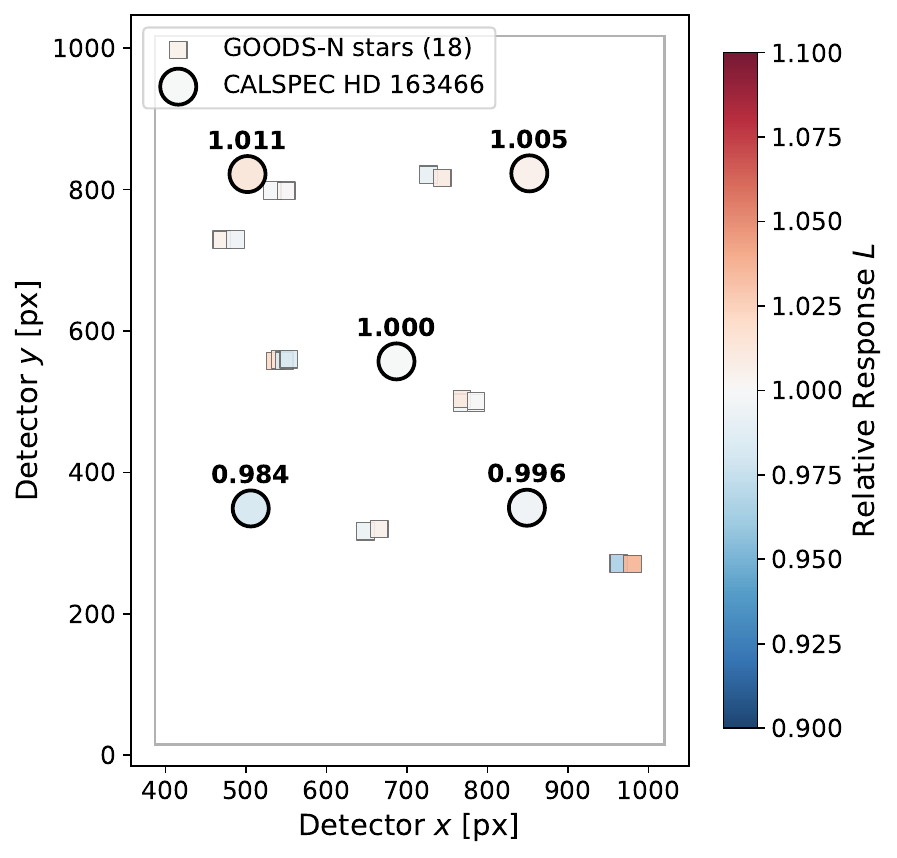}
\caption{
The validation of the L-flat $L(x, y)$ through the response residuals of the CALSPEC grid positions (large circles) and of repeated GOODS-N star spectra (squares) across the WFSS detector region.
The field dependence is consistent with unity (max $|L-1| = 0.016$ on the CALSPEC grid).
\label{fig:lflat}}
\end{figure}

We measure the large-scale field dependence of the response in two independent ways (Figure~\ref{fig:lflat}).
First, with the five-position grid of the CALSPEC standard, we measure the ratio of each response curve to the combined curve $f_R$ at widely separated field positions.
We measure a maximum deviation from unity of 0.016 (MAD 0.003).
Second, we use 18 repeated spectra of the bright stars in the GOODS-N field observed at different field positions, and compare each count-rate measurement with the median of each star's set.
We obtain a per-spectrum dispersion of 0.014.
Most of these repeats span only the $\sim$20 pixel dither offsets, so this test mainly constrains the repeatability and small-scale response variations.
A direct comparison with the combined $f_R$ would probe the full field, but it is limited by the per-star SED scale uncertainty ($\sim$4\%; Figure~\ref{fig:fluxcal}, panel R3).
Both tests are consistent with $L(x, y) = 1$ at the percent level, i.e., a uniform L-flat and negligible field dependence of the response.
More data are needed to fully calibrate and validate the L-flat of MIRI WFSS.

\subsection{Released calibration products}
\label{sec:cal_release}

We compile all components in a single versioned release, \relname.
This release contains the flat field, the master sky with its additive defect map and PCA components, the WFSS region mask, the trace and dispersion coefficient tables, the response table with per-bin anchor provenance, the CALSPEC-direct response measurement, the L-flat, and example calibrated spectra, together with a usage recipe and a SHA-256 manifest.
The recipe for data calibration and spectral extraction includes the following major steps:
(1) divide the rate image by the flat, then subtract the additive defect map and the scaled master sky;
(2) compute the spectral trace $\Delta x(x_0, y_0, \Delta y)$;
(3) invert the dispersion relation $\Delta y(x_0, y_0, \lambda)$ for the wavelength of each row;
(4) sum over the cross-dispersion aperture with local background subtraction;
(5) convert the count rate to flux density by dividing by $f_R(\lambda)$.
The release and an example reduction notebook are hosted in a GitHub repository\footnote{\url{https://github.com/fengwusun/miri_wfss}}.
The release is also archived on Zenodo (DOI: \dataset[10.5281/zenodo.22803722]{https://doi.org/10.5281/zenodo.22803722}).

\begin{figure*}[!t]
\centering
\includegraphics[width=\linewidth]{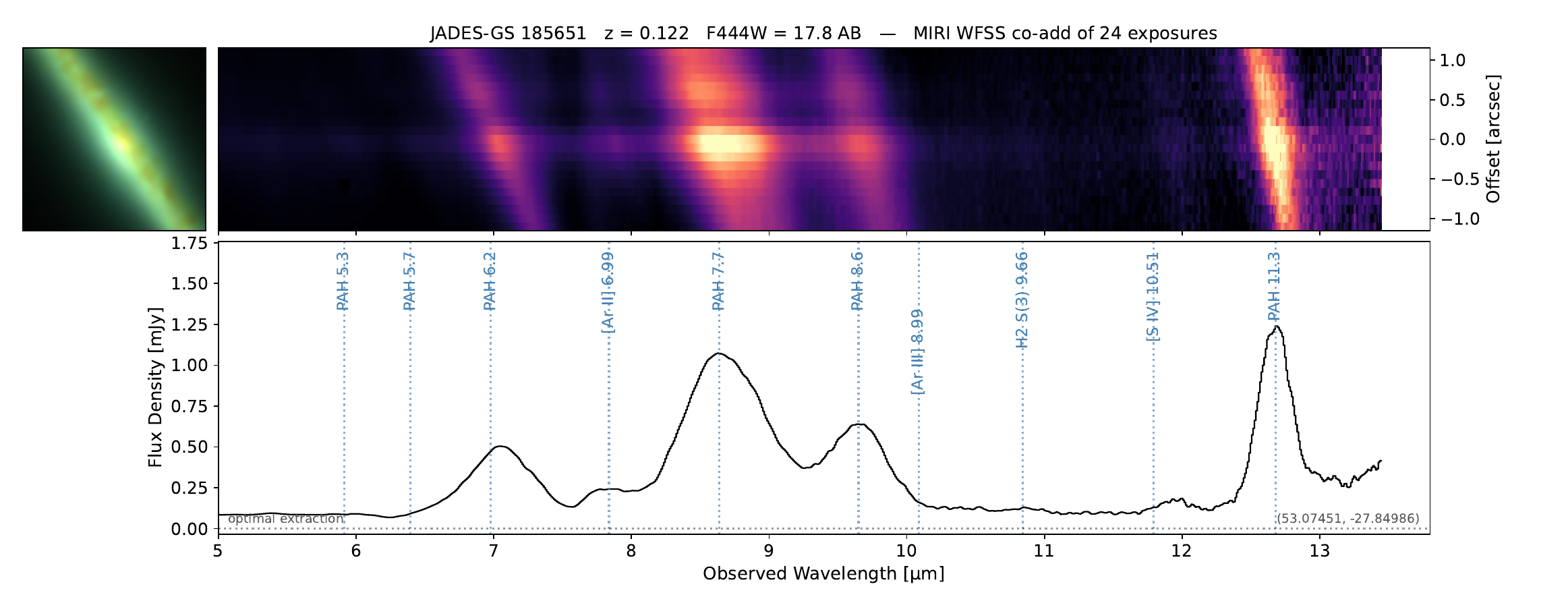}
\includegraphics[width=\linewidth]{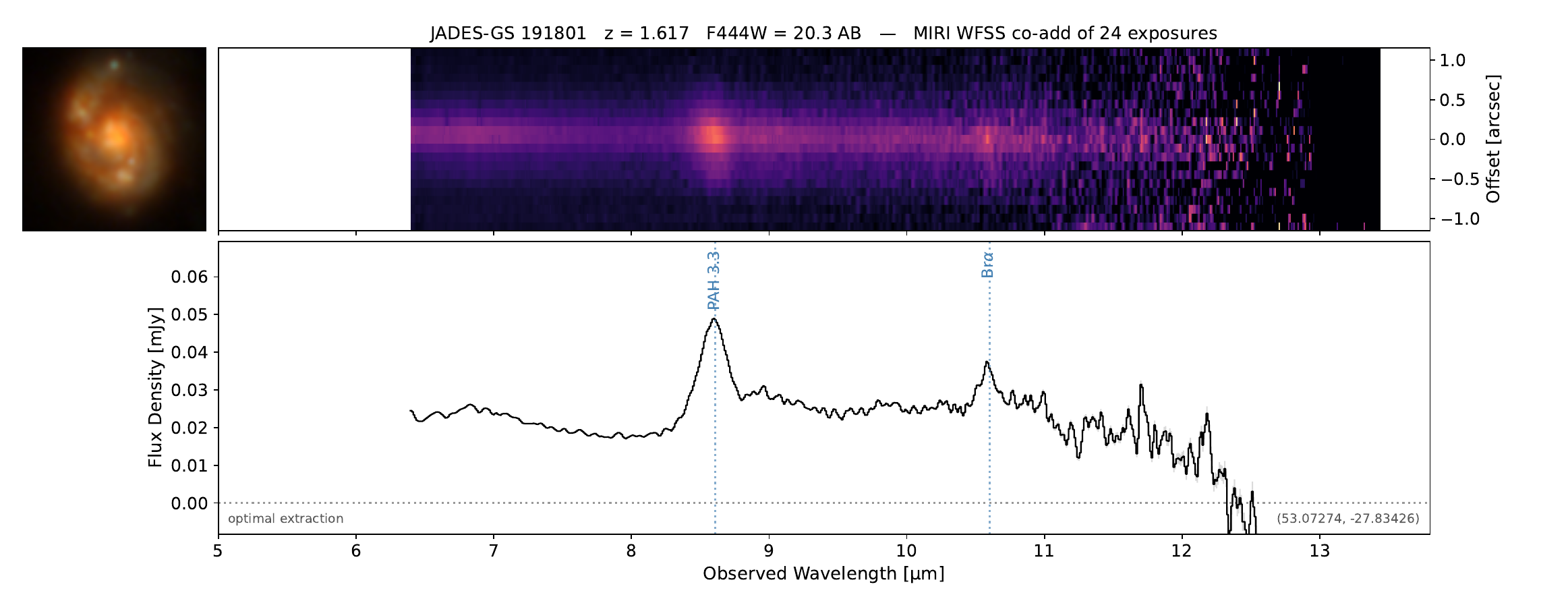}
\includegraphics[width=\linewidth]{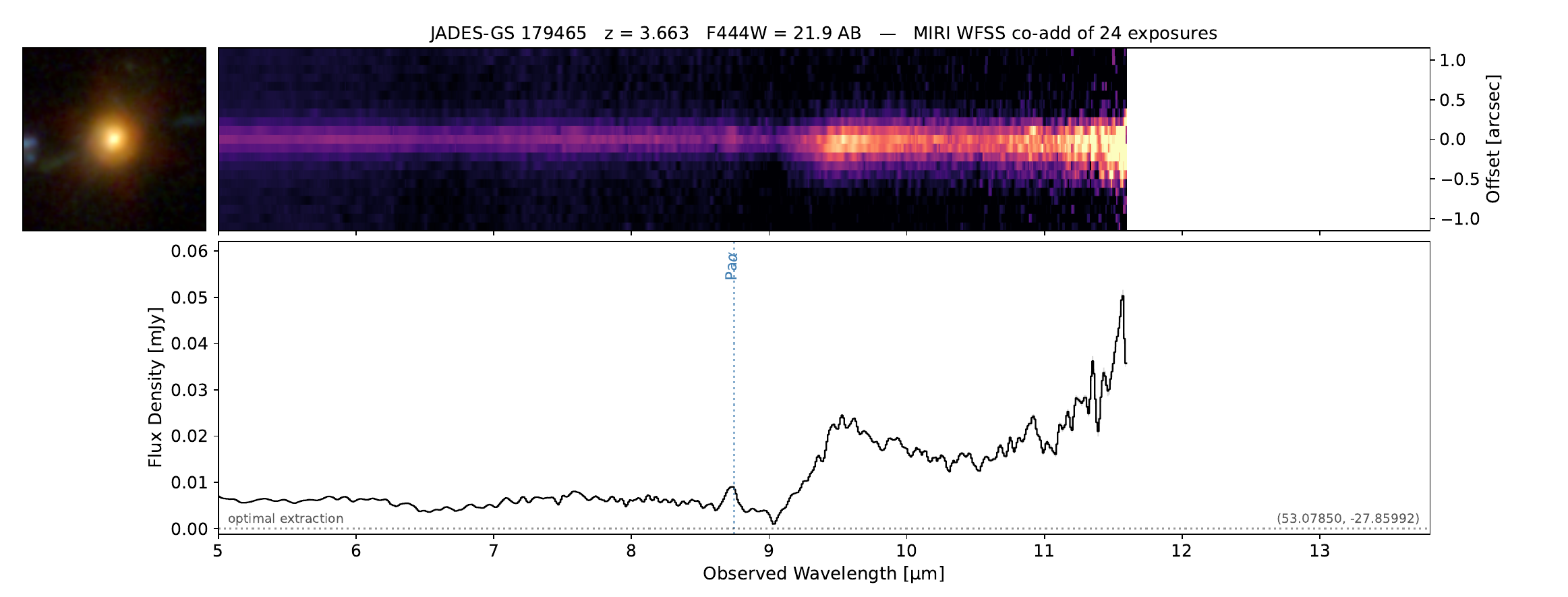}
\caption{
Three examples of MIRI 2D and 1D slitless spectra of bright galaxies.
Top: JADES-GS\,185651 ($z=0.122$), an edge-on disk galaxy with prominent PAH features.
Middle: JADES-GS\,191801 ($z=1.617$) with the PAH 3.3\um\ feature and the \bra\ line detected.
Bottom: JADES-GS\,179465 ($z=3.663$) with the \paa\ line detected (note that the continuum at $>$9\,\micron\ is subject to a contaminant).
For each source, we show its NIRCam F444W--F200W--F115W RGB cutout image (top-left), rotated so the dispersion direction points right and with the same angular height as the 2D spectrum; the flux-calibrated co-added 2D spectrum (top-right); and the optimally extracted 1D spectrum with features marked at the spectroscopic redshift (bottom-right).
\label{fig:examples}}
\end{figure*}

\section{A Mid-Infrared Spectral Atlas of GOODS Galaxies}
\label{sec:atlas}

To demonstrate the quality of our calibration and capability of our MIRI WFSS data reduction pipeline, we construct a mid-infrared spectral atlas of galaxies in the GOODS fields with archival data.
The MIRI/WFSS spectral atlas is visually inspected and released as a product of this work.

\subsection{Sample selection}
\label{sec:atlas_selection}

We select targets by cross-matching the JADES photometric catalogs with the spectroscopic-redshift compilations of each field (Section~\ref{sec:obs_ancillary}) and with the MIRI WFSS footprints (Figure~\ref{fig:footprints}).
We extract the spectra of $\sim$2{,}000 galaxies across the two fields and retain those with median continuum signal-to-noise ratio ${\rm S/N} \ge 1.5$ per 0.01\um\ bin.
We then visually inspect every spectrum, removing spectra with obvious artifacts (e.g., those too close to the detector edges) and spectra with strong contamination from overlapping traces.
Most galaxies are too faint to be detected with MIRI WFSS.
The final atlas contains \Natlas\ galaxies (\NatlasGN\ in GOODS-N, \NatlasGS\ in GOODS-S) at $z = 0.08$--3.71, with NIRCam F444W magnitudes of 17.6--23.7 AB.
We describe the release of the spectral atlas in Appendix~\ref{sec:appendix_table}.

\begin{figure*}[!t]
\centering
\includegraphics[trim={0 12pt 0 0},clip,width=\linewidth]{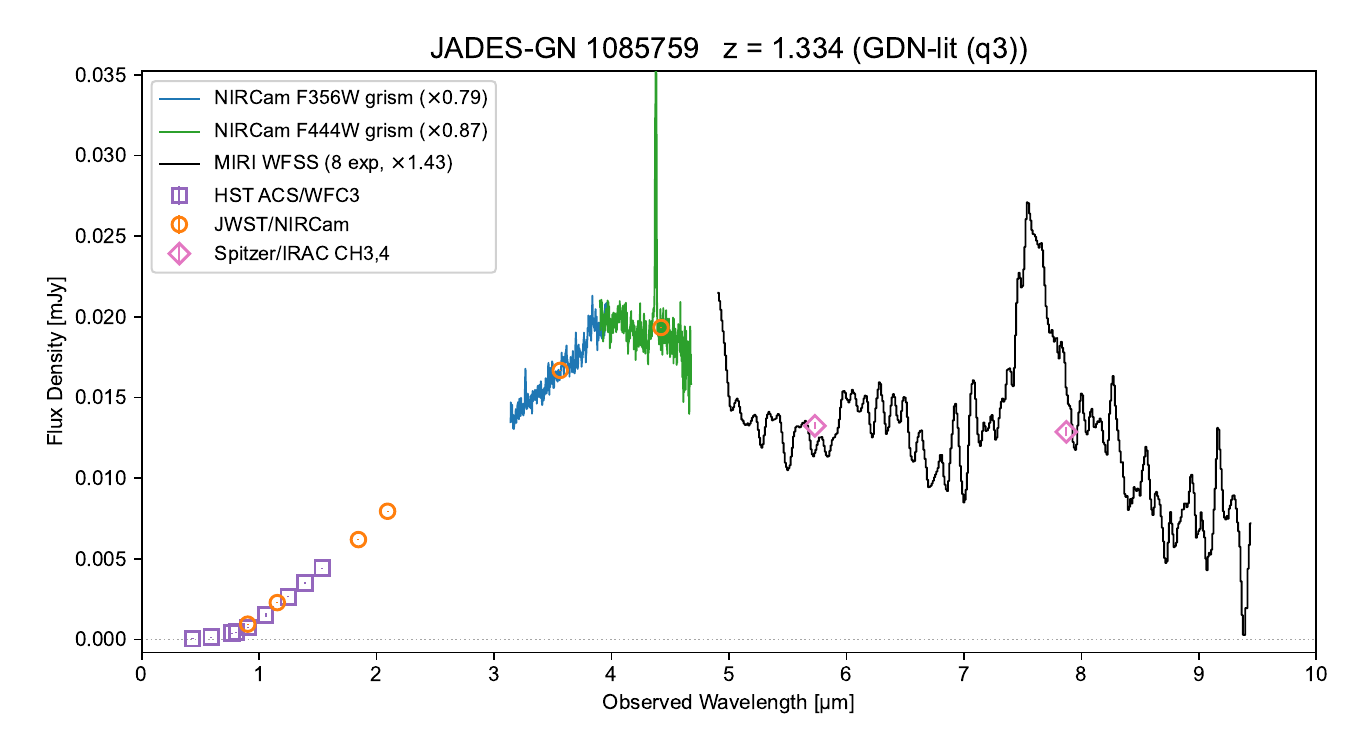}
\caption{
The optical-to-mid-IR SED of JADES-GN\,1085759 at $z=1.334$, including JADES Kron photometry (\hst\ ACS/WFC3, squares; NIRCam, circles), archival \spitzer/IRAC 5.8/8.0\um\ photometry (diamonds), NIRCam grism spectra (\congress\ F356W, blue; \fresco\ F444W, green), and the MIRI WFSS spectrum (black; this work).
Spectra are rescaled onto the photometric SED through synthetic photometry in the matching bands. 
\paa\ (4.38\um) and PAH 3.3\um\ (7.7\um\ observed) are detected.
\label{fig:sed}}
\end{figure*}

\subsection{Spectral extraction and co-addition}
\label{sec:atlas_extraction}

We process each covering exposure following the calibration and extraction procedures of Sections~\ref{sec:obs_processing} and \ref{sec:cal_release}.
We extract the 2D rectified spectrum of each source from each exposure, with proper propagation of the error and data-quality arrays.
After flux calibration, each exposure is resampled to a common 0.01\um\ wavelength grid.
We combine the exposures with an inverse-variance-weighted, sigma-clipped mean.
The combination is restricted to the dominant dispersion position-angle group ($\pm10\arcdeg$), so that the co-added cross-dispersion axis has a single sky orientation.

We apply various treatments to reject bad pixels when the number of exposures is limited.
For two-exposure stacks, we reject the brighter pixel of any pair that disagrees by more than $4\times$ the propagated pair noise, since cosmic-ray hits are positive.
At near-identical dithers, a detector pixel corrupted by a badly corrected jump can land at the same rectified position in most exposures, survive the per-pixel clipping, and carry an underestimated uncertainty that dominates the inverse-variance weights of the optimal extraction \citep{horne1986}.
We therefore reject sharp negative outliers ($<-7\sigma$ below a running median) from each co-added 2D spectrum. 
Flagged pixels (0.11\% of the co-added atlas pixels) are patched through interpolation along the wavelength direction, and their uncertainties are reset to the local median error.
The sampling rate along the spatial direction is the same as the MIRI native pixel scale (0\farcs11/pixel).

We extract 1D spectra from the calibrated, co-added 2D spectra with the optimal algorithm \citep{horne1986}.
The spatial profile is the NIRCam F444W image of the source, rotated to align with and collapsed along the dispersion axis, and recentered on the observed trace centroid.
A small-aperture boxcar extraction ($\pm3$\,pixel\,$=\pm0\farcs33$, with profile-based aperture correction) serves as a fallback. 
We adopt the extraction with the higher continuum S/N; the optimal extraction dominates the sample at a fraction of $\sim$90\%, with a typical $\sim$10--15\% S/N gain.
Although the fluxes extracted through the boxcar and optimal methods generally agree with each other, we caution that they are profile-weighted within the $\pm1\farcs1$ extraction window.
For extended sources they are not total fluxes, and comparisons against large-aperture photometry require an even larger aperture correction.

\subsection{Example spectra}
\label{sec:atlas_examples}

Figure~\ref{fig:examples} shows three representative spectra of our mid-IR spectral atlas.
Each panel shows the NIRCam color image rotated to align with the dispersion frame (with the displayed height exactly matching the 2D cross-dispersion extent), the co-added 2D spectrum, and the optimally extracted 1D spectrum with emission features marked at the spectroscopic redshift.
JADES-GS\,185651, an edge-on disk galaxy at $z=0.122$, displays the full PAH 6.2/7.7/8.6/11.3\um\ sequence spatially resolved along the disk.
At higher redshifts, PAH 3.3\um\ feature and \bra\ line are detected with JADES-GS\,191801 ($z=1.617$), and \paa\ line is detected with JADES-GS\,179465 ($z=3.663$).
We note that the spectrum of JADES-GS\,179465 at $>9$\um\ is subject to a bright continuum contaminant.

\subsection{Synergy with NIRCam grism spectroscopy}
\label{sec:atlas_sed}

For sources in GOODS-N, the \congress\ (F356W) and \fresco\ (F444W) NIRCam grism spectra extend the wavelength coverage blueward to 3.1\um, and marginally overlap the MIRI WFSS at 4.7--5.0\um.
This provides a unique dataset for comparing the spectral line features of bright galaxies at $z\sim1$ across a wide wavelength range (3--14\,\micron) and at high completeness.

Figure~\ref{fig:sed} shows the optical-to-mid-IR SED of JADES-GN\,1085759, a galaxy at $z=1.334$.
The SED includes \hst\ ACS, WFC3-IR, and JWST NIRCam Kron-aperture photometry made available through JADES, as well as archival \spitzer/IRAC 5.8/8.0\um\ photometry \citep{seip2019}.
The three independently calibrated slitless spectra from NIRCam F356W, F444W, and MIRI are overlaid.
We rescale each spectrum onto the photometric SED through synthetic photometry in the matching bands, absorbing the aperture differences among the three extractions. 
\paa\ appears in emission in the F444W grism coverage, and the PAH 3.3\um\ band appears within the MIRI coverage.

\begin{figure*}[!t]
\centering
\includegraphics[trim={12pt 12pt 12pt 0},clip,width=0.7\linewidth]{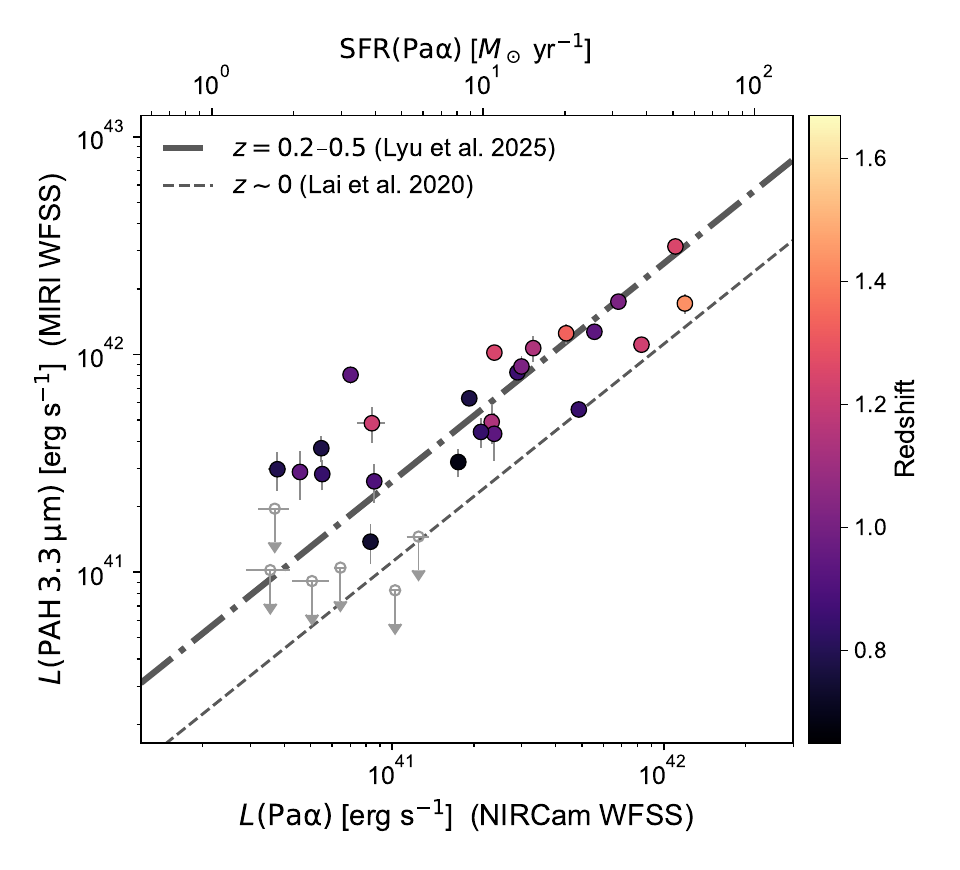}
\caption{
\paa\ luminosities from NIRCam WFSS versus PAH 3.3\um\ luminosities from MIRI WFSS for the 30 GOODS-N galaxies in our mid-IR spectral atlas at $z = 0.65$--1.67.
All sources are color-coded by their redshifts.
Filled circles are PAH 3.3\um\ detections (${\rm S/N} \ge 3$); open circles with arrows are 2$\sigma$ upper limits.
The top axis converts $L({\rm Pa\alpha})$ to SFR assuming Case B recombination ($L_{\rm H\alpha} = 8.61\,L_{\rm Pa\alpha}$) and the H$\alpha$-based SFR calibration of \citet{kennicutt2012}.
The dashed and dash-dotted lines show the SFR--$L_{3.3}$ relations of \citet{lai2020} and \citet{lyuj2025}, respectively, converted through the same \paa-to-SFR calibration.
}
\label{fig:paa_pah33}
\end{figure*}

The joint wavelength coverage enables emission-line and PAH synergy that neither instrument delivers alone at $z\simeq 0.5 - 1.7$.
As a demonstration in Figure~\ref{fig:paa_pah33}, we compare the \paa\ luminosities measured by NIRCam WFSS with the PAH 3.3\um\ luminosities measured from our MIRI WFSS spectra.
These 30 galaxies at $z = 0.65$--1.67 are drawn from our mid-IR spectral atlas, with \paa\ detections (${\rm S/N} \ge 3$) and line-flux measurements from the \congress\ or \fresco\ NIRCam WFSS data \citep{Sun_CONGRESS}.
For each source, we measure the PAH 3.3\um\ band flux from the MIRI spectrum. 
We fit an inverse-variance-weighted linear continuum over the rest-frame 3.05--3.17 and 3.45--3.58\um\ sidebands, and integrate the continuum-subtracted flux over rest-frame 3.20--3.42\um.
The PAH 3.3\um\ band is detected (${\rm S/N} \ge 3$) in 24 of the 30 galaxies.
Luminosities are computed for a flat $\Lambda$CDM cosmology with $H_0 = 70$\,km\,s$^{-1}$\,Mpc$^{-1}$ and $\Omega_{\rm m} = 0.3$.

Figure~\ref{fig:paa_pah33} shows a clear correlation between the two luminosities over a range of $\sim$1.5\,dex, corresponding to an SFR range of $\simeq$2--60\,$M_\odot$\,yr$^{-1}$ assuming Case B recombination and the H$\alpha$-based SFR calibration of \citet{kennicutt2012}.
This spectroscopic result suggests that PAH\,3.3\,\micron\ remains a valid SFR tracer, extending from the local Universe \citep{lai2020} and $z\sim0.4$ \citep{lyuj2025} out to $z\sim1.6$.
We measure a median ratio of $L({\rm PAH\,3.3})/L({\rm Pa\alpha}) = 2.8$ for galaxies in this sample. 
We find that the $L({\rm PAH\,3.3})$--SFR relation of these $z\sim1$ galaxies is consistent with that of galaxies at $z\simeq0.2$--0.5 as measured by \citet{lyuj2025}, while the $L({\rm PAH\,3.3})/{\rm SFR}$ ratio is higher than that of luminous infrared galaxies in the local Universe measured through AKARI \citep{lai2020}.

We caution that neither luminosity is corrected for dust attenuation.
Assuming a galaxy-wide attenuation of $A_V = 1$\,mag and the Milky Way extinction curve of \citet{gordon2023}, the corrections would be 13\% for \paa\ and 5\% for PAH 3.3\um, and the dust-corrected $L({\rm PAH\,3.3})/L({\rm Pa\alpha})$ ratio would decrease by 8\%.
We also caution that the PAH and \paa\ spatial profiles can differ from the spectral extraction profile constructed from the F444W image (Section~\ref{sec:atlas_extraction}), and therefore the absolute ratio carries a systematic uncertainty for extended sources.
Joint measurements of hydrogen recombination lines and PAH bands for hundreds of galaxies are a unique product of the overlapping NIRCam and MIRI WFSS coverage in the GOODS fields.

\begin{figure*}[!t]
\centering
\includegraphics[width=\linewidth]{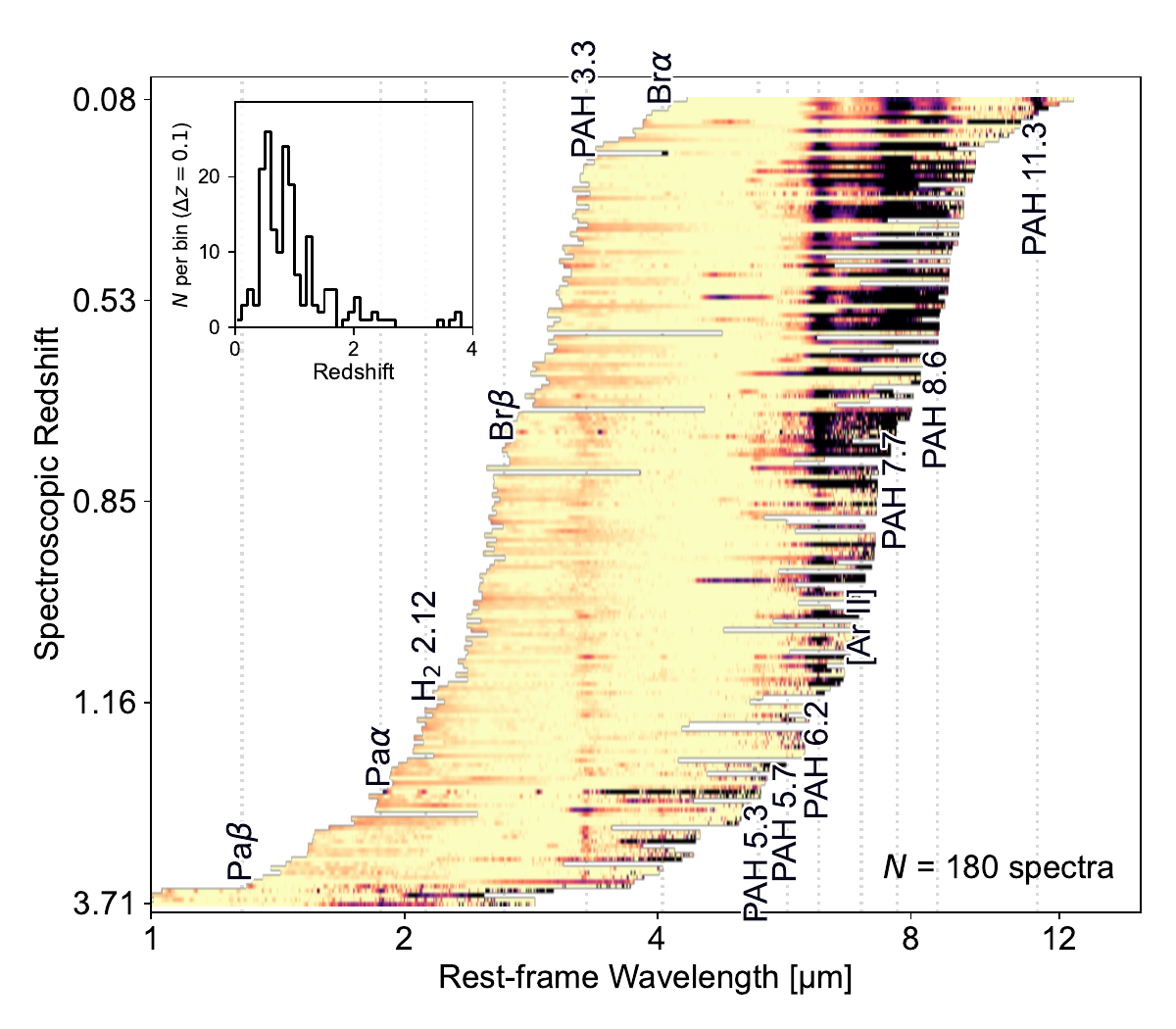}
\caption{
The MIRI WFSS spectral atlas of \Natlas\ galaxies in the GOODS-N/S fields, shown in the rest frame.
Each row is one galaxy, sorted by redshift. The redshifts of the sample at the 0/25/50/75/100th percentiles are shown on the y-axis.
All spectra are normalized by the sources' F444W flux densities and continuum-subtracted.
Major PAH bands and recombination lines are labeled with vertical dotted lines.
An inset panel (top-left) shows the redshift histogram.
\label{fig:stack2d}}
\end{figure*}

\subsection{Stacking analyses}
\label{sec:atlas_stack}

To characterize the typical mid-IR spectrum of the atlas, we project each 1D spectrum to the rest frame, normalize it by the source's F444W flux density, and resample it onto a common logarithmic rest-wavelength grid.
In Figure~\ref{fig:stack2d}, we subtract each spectrum's continuum with a sigma-clipped median and display the spectral atlas as a redshift-sorted ``spectral ladder''.
The PAH bands appear as coherent vertical structures across the sample.
At the high-redshift end, \paa\ is robustly detected in deep GO-8544 WFSS data for galaxies out to $z=3.71$.

\begin{figure*}[!t]
\centering
\includegraphics[width=\linewidth]{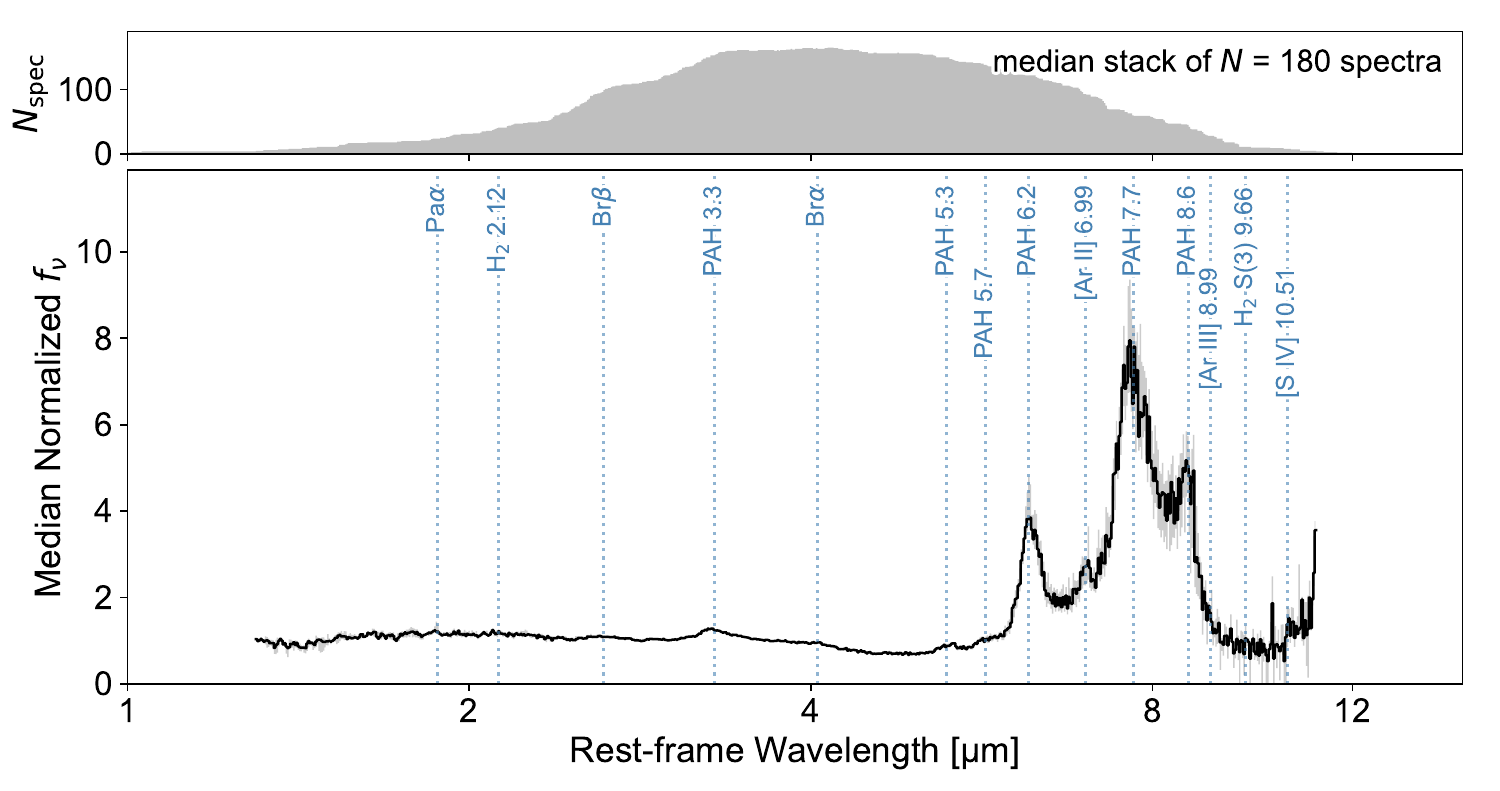}
\caption{
Median rest-frame stack of the \Natlas\ mid-infrared spectra in our atlas (bottom) and the number of contributing spectra per wavelength bin (top).
Each spectrum is divided by its sigma-clipped median continuum level before stacking, so the composite retains the continuum shape.
The PAH 3.3, 5.3, 6.2, and 7.7\um\ bands are detected, and \paa\ is tentatively recovered.
The gray band is the per-bin 16--84th percentile bootstrap uncertainty of the median stack spectrum.
\label{fig:stack1d}}
\end{figure*}

Figure~\ref{fig:stack1d} shows the median stack of all \Natlas\ spectra, revealing a textbook star-forming composite at near- to mid-IR wavelengths.
Here we instead divide each spectrum by its sigma-clipped median continuum, so the composite retains the continuum shape.
We measure feature significances as the peak minus the local continuum, in units of the bootstrap uncertainty of the median stack.
The detected features are PAH 3.3\um, 6.2\um, 7.7\um\ complex, and 5.3\um.
The stack ends at rest-frame $\simeq$11\um, where fewer than five spectra overlap, and the PAH 11.3\um\ band is therefore only visible in the lowest-redshift rows of Figure~\ref{fig:stack2d}.
\paa, PAH 5.7\um, \brb, and PAH 8.6\um\ features are weak ($\simeq$1.5--2.5$\sigma$) in the stack.
\bra, the \htwo\ rotational lines, \arii, and \siv\ remain undetected in the median stack, though detected in certain bright sources.

We caution that the composite is illustrative rather than a population measurement.
Across our spectral atlas sample, the rest-wavelength coverage of each source strongly depends on its redshift, and the median stacking and normalization give each galaxy equal weight regardless of its brightness or luminosity.

\subsection{Cross-instrument comparison: GN-IRS-15}
\label{sec:atlas_xinstr}

\begin{figure*}[!t]
\centering
\includegraphics[width=\linewidth]{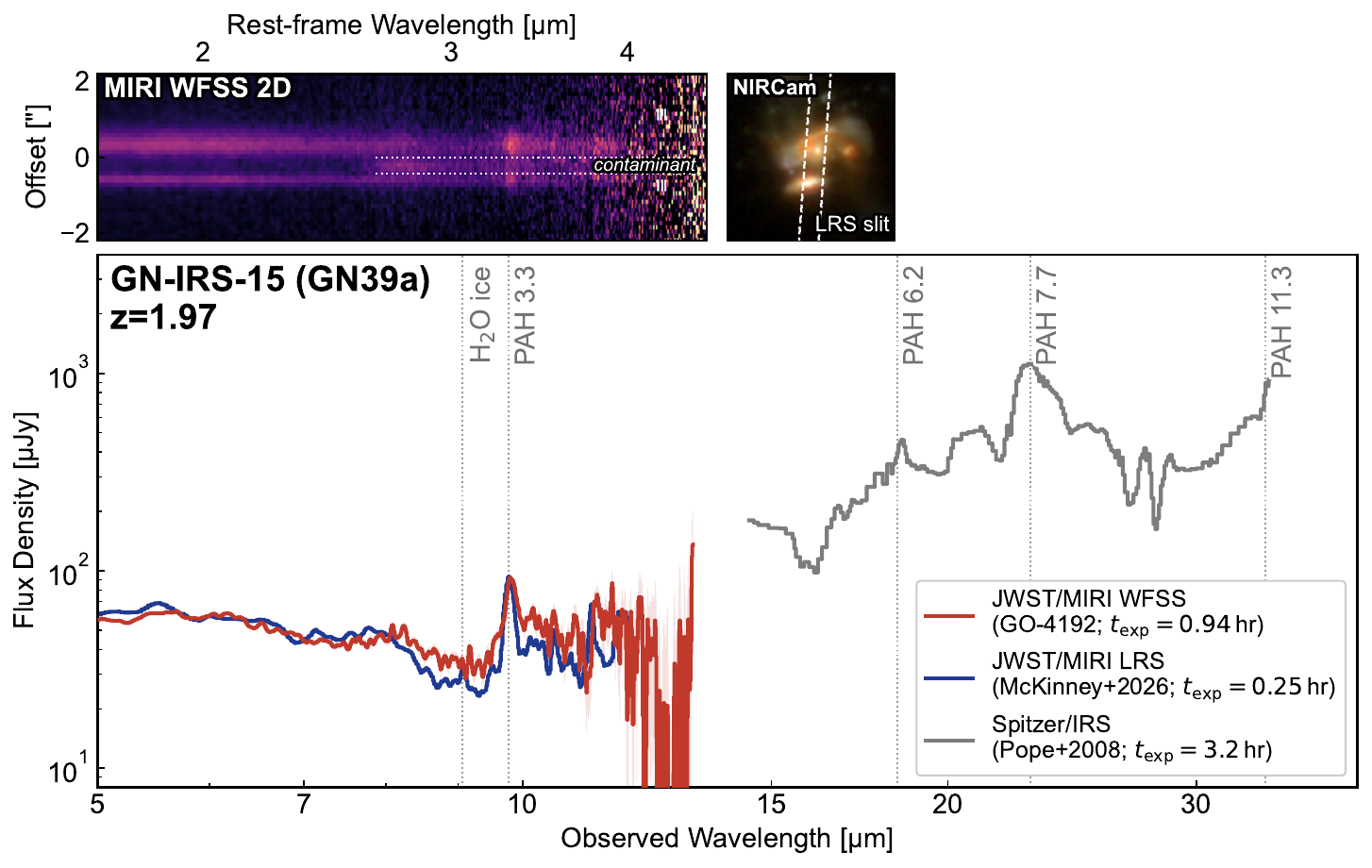}
\caption{GN-IRS-15 ($=$GN39a of \citealt{pope2008}; JADES-GN 1082021), a luminous infrared galaxy at $z=1.97$ observed by three mid-infrared spectrographs.
\textbf{Top left:} co-added MIRI WFSS 2D spectrum, aligned with the observed-wavelength axis of the panel below.
\textbf{Top right}: the NIRCam F444W--F200W--F115W RGB cutout ($4\farcs4\times4\farcs4$) of the galaxy, rotated so the MIRI dispersion direction points right, with the same cross-dispersion extent as that of the 2D spectrum.
The white dashed rectangle marks the combined MIRI LRS slit of GO-3224.
\textbf{Bottom}: the MIRI WFSS 1D spectrum, 
the \jwst/MIRI LRS spectrum of the same galaxy from \citet[][fixed slit]{mckinney2026}, and the archival \spitzer/IRS spectrum from \citet{pope2008}.
The per-instrument on-source exposure times are labeled.
The dotted lines in the 2D panel mark the lane of the contaminating trace of JADES-GN 1082239, excluded from the 1D extraction (Section~\ref{sec:atlas_xinstr}).
The three observations together trace the rest-frame $\simeq$1.7--13\um\ spectrum, including the water ice 3\um\ absorption and the 3.3, 6.2, 7.7, and 11.3\um\ PAH features (dotted lines).
\label{fig:threeway}}
\end{figure*}

GN-IRS-15 \citep[$=$GN39a;][]{pope2008}, one galaxy in our spectral atlas, was observed with \jwst/MIRI LRS by \citet{mckinney2026}.
It was also observed by \spitzer/IRS \citep{pope2008}, and a comparison of the mid-IR spectra taken with three instruments/modes is shown in Figure~\ref{fig:threeway}.
GN-IRS-15 is a luminous infrared galaxy and a complex merger at $z=1.97$ \citep{mckinney2026}, also identified as JADES-GN 1082021 (${\rm R.A.} = 189.29736\arcdeg$, ${\rm Decl.} = +62.22534\arcdeg$).

The 2D spectrum resolves the two components of the merger, and it also reveals a third, contaminating trace in the lane between them redward of 7.9\um.
We identify the contaminant as JADES-GN 1082239 ($m_{\rm F444W} = 22.4$), a galaxy $7\farcs5$ away along the dispersion direction whose cross-dispersion offset ($-0\farcs25$) matches the lane, and whose blue spectral cutoff at 4.7\um\ lands exactly at the observed 7.9\um\ onset of the extra light.

We therefore extract the WFSS 1D spectrum through two boxes of 6 and 3 pixels placed on the merger components, with the geometry optimized for continuum S/N while suppressing the contaminant leakage.
We correct the aperture losses with the F444W light profile convolved with the wavelength-dependent \texttt{STPSF} MIRI PSF, and the correction factor is 1.80--1.95 over 5.6--12.8\um.
The corrected spectrum reproduces the archival IRAC 5.8\um\ photometry of the blended system \citep{wang2010} to 3\%.
The MIRI/LRS slit spectrum of \citet{mckinney2026} is rescaled to the IRAC 5.8 and 8.0\um\ photometry to correct for slit losses, and the two spectra indeed agree to $\simeq$6\% over 5--8\um, including the 3\um\ water-ice absorption and the 3.3\um\ PAH feature.
Redward of 8\um\ the WFSS spectrum is $\simeq$25\% brighter than that of the LRS slit spectrum. 
At longer wavelengths, the \spitzer/IRS spectrum extends the coverage through the 6.2, 7.7, and 11.3\um\ PAH complexes \citep{pope2008}.
The LRS and IRS observations each required a dedicated pointing on this single target, whereas the WFSS spectrum was taken together with every other source in the field.
The agreement across three instruments and observatories independently validates our trace, wavelength, and flux calibration, and shows that the extracted spectra are ready for science.

%

\section{Caveats and Suggestions for Future MIRI WFSS Observations}
\label{sec:caveats}

The archival data analyzed in this work effectively constitute an unplanned MIRI WFSS survey.
Their heterogeneity exposes several practical lessons for future MIRI WFSS observation planning and calibration reuse.
We collect and present these lessons in this section.

We first note that source contamination and spectral overlap are caveats common to all WFSS observations, and they apply to the MIRI WFSS data as well.
We advise caution when interpreting spectral features detected in the 2D and 1D spectra.
The rest of this section presents caveats and suggestions on sensitivity, the sky background and its subtraction, and astrometry.

\begin{figure*}[!t]
\centering
\includegraphics[width=\linewidth]{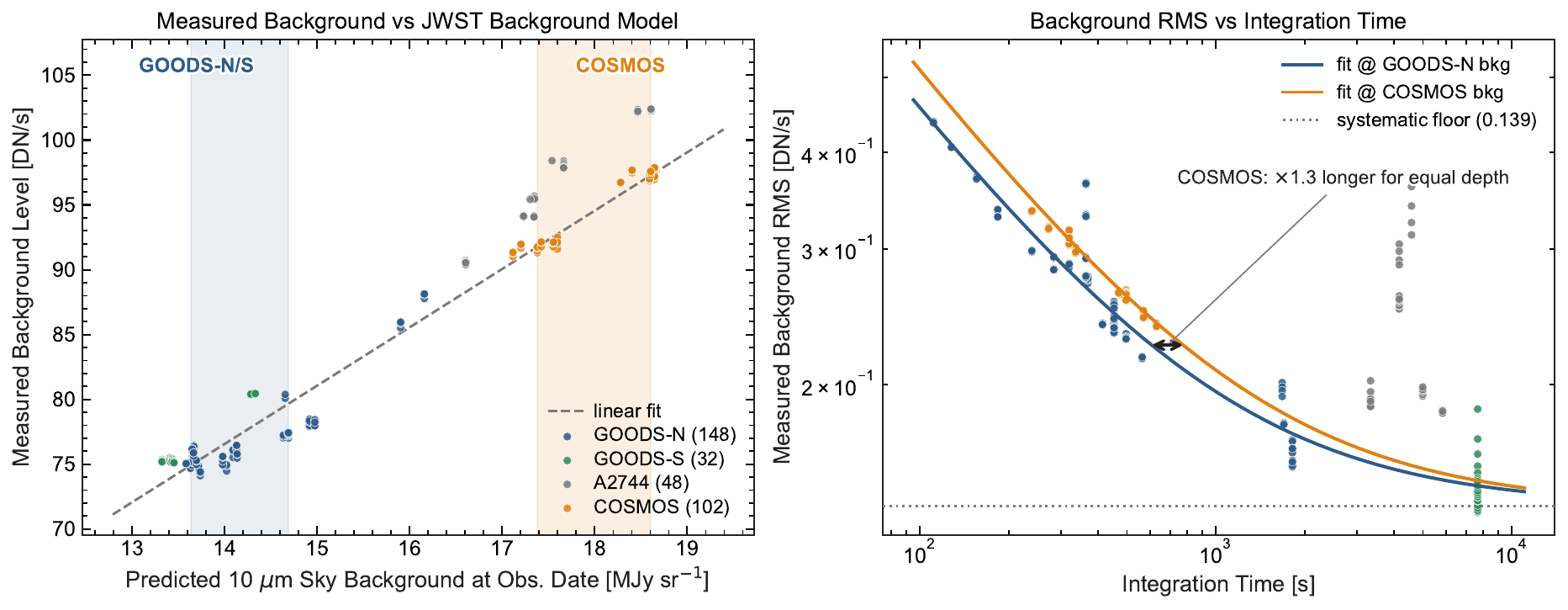}
\caption{
Empirical field-to-field background comparison from the archival P750L exposures.
\textbf{Left}: the measured background level of each exposure against the total 10\um\ sky background predicted by the JWST background tool at its observation date.
The shaded bands mark the predicted-background range of the GOODS-N/S frames versus the higher-background COSMOS frames (GO-4192, $\beta=-9\arcdeg$; $\sim$34\% above GOODS-N as flown, $\sim$53\% on a median day), with the Abell 2744 frames of GO-8051 ($\beta=-29\arcdeg$) in between.
\textbf{Right}: measured background RMS in the source-masked, calibrated WFSS region versus integration time, fit with the background-limited noise model of Section~\ref{sec:caveats_depth}.
The solid curves show the fits at the median GOODS-N and COSMOS backgrounds.
$1.3\times$ longer integration is required for COSMOS to achieve equal depth at the as-flown background.
The deep ($\gtrsim$2 ks) exposure frames flatten onto the systematic floor.
\label{fig:fieldbkg}}
\end{figure*}

\subsection{Sensitivity and the noise floor}
\label{sec:caveats_depth}

In calibrated exposures, we find that the source-masked residual background RMS exceeds the propagated pixel noise in the error extension (ERR), and the excess grows with integration time.
The RMS/ERR ratio is 1.25--1.33 for the $\sim$450\,s GOODS-N tiling exposures, $\simeq$1.9 for the 1.7\,ks GO-4762 frames, and $\simeq$3.1 for the very deep 7.7\,ks GO-8544 frames.
We note that as of this writing, incorrect pipeline error estimates are a known issue for MIRI LRS data products (MIRI-LRS05)\footnote{\url{https://jwst-docs.stsci.edu/known-issues/miri-known-issues/miri-lrs-known-issues}}.
This issue may contribute to the overall normalization of the RMS/ERR ratios, but the growth of the excess with integration time is independent of the ERR scale.

The right panel of Figure~\ref{fig:fieldbkg} demonstrates this behavior.
We fit the measured background RMS of the GOODS-N, GOODS-S, and COSMOS frames with a background-limited noise model, $\sigma = (C\,B/t + \sigma_{\rm floor}^2)^{1/2}$, where $B$ is the per-frame background level, $t$ is the integration time, and $C$ is a scaling constant.
Freeing the power index of the exposure-time term gives $-0.94 \pm 0.02$, consistent with the photon-limited $t^{-1}$ scaling adopted here.
The $\gtrsim$2\,ks MIRI WFSS exposures flatten onto the fitted systematic floor of $\sigma_{\rm floor} = 0.139$\,\dnps.
In other words, the RMS noise of single integrations beyond $\sim$1--2\,ks is no longer photon-limited.
We attribute the dominant error budget to (\romannumeral1) a floor of calibration systematics (e.g., flat and sky-model residuals) and, in the deepest data, (\romannumeral2) genuine confusion from unmasked faint sources.
The astrophysical part of this error floor does not average down with repeated exposures at a fixed pointing.
We also note that the floor is not reduced by rebuilding the flat-field and sky-background calibration with the improved MIRI imaging flats delivered recently (e.g., \texttt{jwst\_miri\_flat\_0844}).
Any static flat-field error is absorbed into the master-sky template and cancels in the sky subtraction, and therefore the choice of the flat-field version has negligible impact on the depth of MIRI WFSS data.

\begin{figure*}[!t]
\centering
\includegraphics[width=0.75\linewidth]{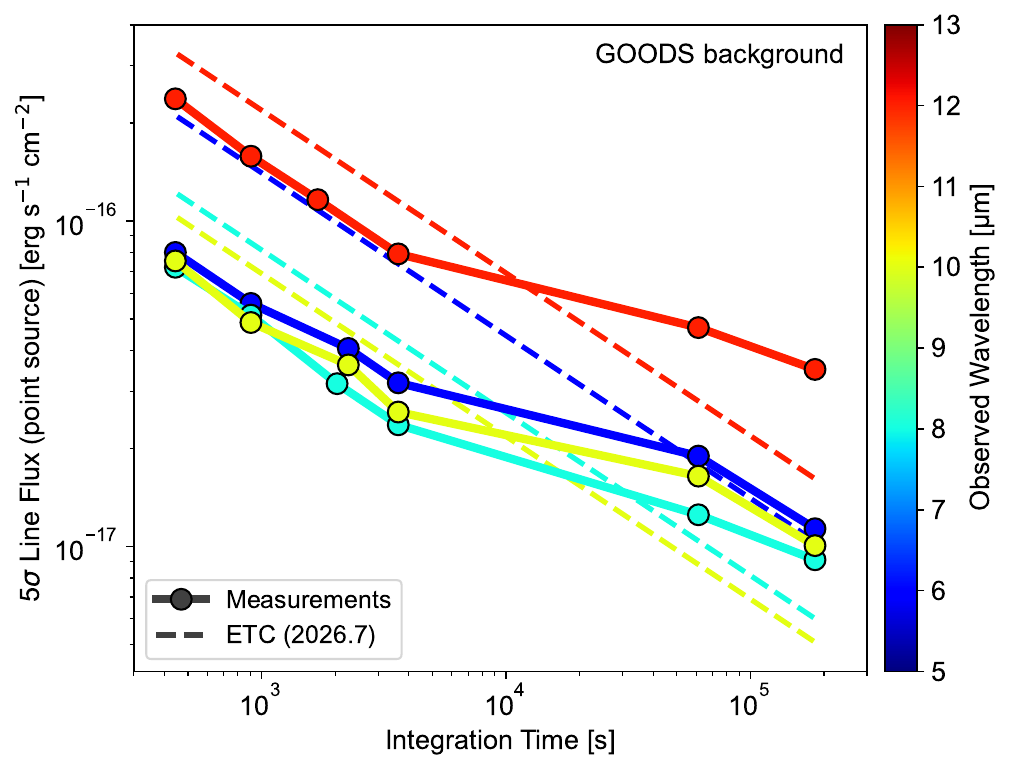}
\caption{$5\sigma$ emission-line sensitivity of MIRI WFSS versus integration time under the typical GOODS-field background ($\sim$76\,\dnps), color-coded at observed wavelengths of 6, 8, 10, and 12\um.
The line sensitivities are measured by injecting unresolved emission lines of point sources into 1D spectra, which are force-extracted at blank sky positions in the real GOODS-N/S data, with a boxcar extraction height of 5\,pixel (0\farcs55).
Dashed lines show the predictions of the JWST ETC (\texttt{Pandeia} version 2026.7) under the same background, extraction aperture, and line profiles.
The measured line sensitivity is $\sim1.8$ times deeper than the ETC prediction at $t \lesssim 4$\,ks.
However, beyond $\sim$10\,ks, the measured sensitivity curves flatten and therefore the photon-limited ($\sigma \propto t^{-1/2}$) sensitivity curve with the ETC becomes too optimistic.
\label{fig:linesens_time}}
\end{figure*}

Figure~\ref{fig:linesens_time} translates the systematic floor into the emission-line sensitivity of co-added MIRI WFSS spectra. 
The line sensitivity is measured by injecting unresolved emission lines into 1D spectra force-extracted from blank positions in the real GOODS-S/N exposures.
In this injection experiment, we use the line spread function following \citet{kendrew2015}, and the 1D spectra are extracted with $H=5$\,pixel (0\farcs55) boxcar.
Each injected emission line is fitted with Gaussian profile through least squares, and we use the line flux errors measured from the covariance matrix of secure line detections ($\mathrm{S/N} \geq 8 $) to measure the emission-line sensitivity.
The achieved $5\sigma$ line sensitivity follows the photon-limited $t^{-1/2}$ scaling only up to a few ks, and flattens toward the deepest available stacks ($\sim$180\,ks).

We also compare the measured line sensitivity with \texttt{Pandeia} \citep{pandeia} version 2026.7, i.e., the backend of JWST Cycle-6 Exposure Time Calculator (ETC).
With the same observing condition (i.e., background, point source, unresolved line and $H=5$\,pixel extraction aperture), we simulate the $5\sigma$ emission line sensitivities.
With the deepest exposure ($\sim$180\,ks), the JWST ETC over-predicts the achievable depth by up to $\sim$50\%.
However, at $t \lesssim 4$\,ks, the achieved sensitivity is instead $\sim1.8$ times deeper than the ETC prediction.

We therefore recommend building depth from many dithered integrations of $\lesssim$1\,ks per exposure, rather than long integrations at a single dither position.
At the current state of our calibration, per-pointing investments beyond $\sim$10\,ks yield strongly diminishing returns for MIRI WFSS sensitivity.
This is a useful benchmark when weighing one deep pointing against wider tiling.

\subsection{Field selection and the sky background}
\label{sec:caveats_background}

Zodiacal light dominates the 5--14\um\ background with JWST/MIRI, contributing 70--80\% of the total at 5--10\um\ per the STScI background model (\texttt{jwst\_backgrounds}\footnote{\url{https://github.com/spacetelescope/jwst-backgrounds}}; \citealt{rigby23b}).
Ecliptic latitude therefore drives large field-to-field sensitivity differences.
For the median observable day, the model predicts 53\% and 51\% brighter 10\um\ backgrounds in COSMOS ($\beta = -9\arcdeg$) and UDS ($-18\arcdeg$) than in GOODS-N ($+57\arcdeg$).
Low-ecliptic-latitude fields also suffer a $\sim$2$\times$ seasonal background swing (versus $\sim$1.2$\times$ at high latitude) and are observable for much less days of the year than high-latitude fields.

Our archival data confirm such an effect empirically (Figure~\ref{fig:fieldbkg}).
The GO-4192 P750L exposures on COSMOS measure a background 28\% above the GOODS-N/GOODS-S level, and a background RMS 11\% higher at fixed integration time (consistent with the $\sqrt{1.28} \simeq 1.13$ expectation for background-limited data).
The background difference is less than the median-day model prediction because those COSMOS visits happened to execute near the seasonal background minimum.

As an independent check at intermediate latitude, we processed the 48 deep GO-8051 (PI: \'Alvarez-M\'arquez) P750L exposures on Abell 2744 ($\beta = -29\arcdeg$; observed 2025 October--November) through the same calibration pipeline.
The A2744 background is 25\% above GOODS-N. 
With integrations of 4--8\,ks, the A2744 and GOODS-S GO-8544 frames sit on the systematic floor of Section~\ref{sec:caveats_depth}, $\simeq$2.5$\times$ above the extrapolated photon-noise expectation.
Crowding in the A2744 cluster field further inflates its masked-residual RMS, and therefore we present the A2744 data in Figure~\ref{fig:fieldbkg} but exclude them from the fit.

The right panel of Figure~\ref{fig:fieldbkg} demonstrates the penalty in scheduling currency.
At the realized background level, a COSMOS pointing requires $1.3\times$ the integration time of a GOODS-N pointing to reach equal continuum depth, and $1.5\times$ under median-day scheduling.
In other words, one fifth to one third of the time invested at low ecliptic latitude buys no depth at all.

We have two suggestions for future blank-field MIRI WFSS surveys to optimize telescope efficiency:
\begin{enumerate}
    \item We prefer high-ecliptic-latitude fields for deep MIRI WFSS survey programs.
    If a low-latitude field is scientifically required, we suggest constraining the scheduling window toward the zodiacal minimum, at real cost to schedulability.
    
    This suggestion is directly relevant to the first dedicated MIRI WFSS science program.
    In Cycle 5, GO-11605 (PI: Mehta) will invest 43\,hr of P750L wide-field spectroscopy on two NIRISS WFSS fields in COSMOS. 
    The field choice is well motivated by the ancillary NIRISS spectroscopy.
    However, the zodiacal economics above imply that the on-source integration must be $\sim$30--50\% longer to compensate for the brighter foreground. 
    Constraining the visits toward the annual background minimum may recover part of the difference.
    We encourage explicit background-driven scheduling constraints in future low-latitude MIRI WFSS programs.

    \item We also highlight a viable option to use the MIRI LRS slit simultaneously with the MIRI WFSS observations, as in the archival data sets analyzed in this work.
    When a suitable slit target is available, MIRI WFSS observations can be obtained as targeted MIRI LRS observations with the FULL-array readout.

    As demonstrated in Figure~\ref{fig:mode}, MIRI LRS slit benefits from a $\sim$40$\times$ lower sky background, and thus a 5--10$\times$ gain in sensitivity over MIRI WFSS.
    Proper target acquisition before the MIRI LRS observations, together with the accompanying verification images, also enables accurate astrometric verification, and therefore reliable spectral tracing and wavelength calibration.
    Designing MIRI WFSS observations through the MIRI LRS templates, with a target placed onto the LRS slit, can therefore maximize the efficiency of telescope usage.

\end{enumerate}

\subsection{Sky-subtraction mode}
\label{sec:caveats_mode}

The default mode-A subtraction (scaled master sky plus the per-row clipped-median removal) is linear, robust, and sufficient for the data taken in GOODS fields.
The optional mode B adds a five-component PCA correction fitted by linear least squares.
Through our tests, the use of mode B improves the source-masked residual MAD by 2--3\% in GOODS-N frames, 4--10\% in the deep GOODS-S frames, and 14\% on COSMOS compared with mode A.
The COSMOS gain is largest because the master-sky template, mostly dominated by the data in the GOODS fields, does not match the brighter COSMOS zodiacal spectrum perfectly.

However, we note that the PCA sky bases carry a subtle failure mode.
Any compact source signal surviving the per-frame masking is absorbed into the components, and then imprinted into {every} corrected frame at fixed detector positions.
In the released PCA basis, we therefore patch compact $>$4$\sigma$ outliers in each component with the local smooth value.
Difference images of deep GOODS-S and COSMOS frames confirm that the cleaned basis adds no localized features while retaining its correcting power.
However, this cleaning is not perfect, and imprints below the patching threshold can still survive in the released basis at low surface brightness.

Therefore, we recommend mode A as the default background subtraction method, and mode B for low-ecliptic-latitude fields or whenever the residual background retains visible template-shaped structure.
As a general practice, we suggest differencing any PCA-corrected frame against its mode-A counterpart before trusting any compact faint emission.

\subsection{Astrometric accuracy, target acquisition, and the wavelength zero-point}
\label{sec:caveats_astrometry}

Slitless spectral extraction computes the trace and dispersion solutions at the source position predicted by the exposure WCS.
Any error in the actual pointing therefore propagates directly into the extraction, whether from a failed target acquisition (TA), guide-star identification problems, or ordinary visit-level astrometric uncertainty.
This will eventually lead to displacement of the spectral trace along or across the dispersion direction, and therefore errors in wavelength or flux measurements.

Such an uncertainty is not hypothetical; instead, such cases can be found in archival MIRI LRS observations.
For example, one third of the GO-8544 observations suffered a failed target acquisition, caused by an anomalous cosmic-ray event during the TA procedure \citep{helton2025}.
For testing purposes, we measure the actual WFSS astrometry of GO-8544 directly.
We fit trace centroids for seven bright GOODS-S sources across all 32 exposures, and compare them against a GOODS-N control sample.
The per-visit common position offsets are $\le$0.11\,pixel and the common trace tilts are $\le$0.03$\arcdeg$, indistinguishable from the control.
Although the slit TA failure did not corrupt the wide-field WCS, as the WCS is set by the guide-star solution rather than the TA, we highlight the necessity of astrometric verification under such circumstances.
We suggest users verify the per-visit astrometry of any MIRI WFSS data set before trusting absolute wavelengths at the resolution-element level.
This can be done through the accompanying verification images if they exist, or through the trace positions of bright objects with known coordinates.




\section{Summary}
\label{sec:summary}

We present a complete and precise calibration of \jwst/MIRI wide-field slitless spectroscopy, the configuration in which the P750L prism disperses all sources entering the full $\sim70''\times110''$ MIRI imager FoV (Figure~\ref{fig:mode}).
We describe the methodology of the calibration and demonstrate the performance through a mid-IR spectral atlas of galaxies at $z=0$--$4$ in the GOODS fields.
The main results are listed as follows:

\begin{enumerate}
\item We build the flat-field and sky-background calibration. 
The F560W imaging flat performs best among the candidate flats, and the sky is removed with an archival master-sky template, an additive defect map, and an optional five-component PCA refinement (Figure~\ref{fig:lv15seq}).

\item Using spectral traces of Gaia stars in the LMC and bright galaxies and stars in the GOODS fields, we obtain the spectral tracing function for MIRI WFSS to an accuracy of MAD $=$ \tracemad\ pixel (Figure~\ref{fig:trace}).

\item The wavelength solution, anchored on a planetary nebula in the LMC and emission lines from galaxies in the GOODS-N field with known spectroscopic redshifts, is accurate to \wavecalresel\ resolution elements (87--144\,km\,s$^{-1}$) over 7.9--12.8\um\ (Figure~\ref{fig:pn_calibrated}).

\item The flux response is anchored on the CALSPEC standard star \hdstar\ at 7.4--13.5\um\ (absolute accuracy 1--2\%) and extended to 4.7\um\ with an ensemble of bright stars in the GOODS-N field (accuracy $\sim$5\%; Figure~\ref{fig:fluxcal}).
The field-dependent response (L-flat) is unity with an estimated accuracy of $\sim$1\% (Figure~\ref{fig:lflat}).

\item We publicly release the calibration products (\relname) and the reduction pipeline.
This release includes the reference files for every component used by the reduction pipeline, the reduction scripts, and an example notebook that reproduces the spectral extraction on public GOODS-N exposures.

\item Applying the calibration to the archival GOODS data, we construct a visually vetted mid-IR spectral atlas of \Natlas\ galaxies at $z=0.08$--3.71, released with the compiled 2D and 1D spectra (Appendix~\ref{sec:appendix_table}).
PAH bands and atomic lines are detected in individual sources, and the rest-frame median stack detects the PAH 3.3, 5.3, 6.2, and 7.7\um\ bands (Figures~\ref{fig:stack2d} and \ref{fig:stack1d}).
As a science demonstration, we combine the atlas with NIRCam grism spectroscopy into continuous 3.1--13.5\um\ slitless spectra (Figure~\ref{fig:sed}).
The \paa\ and PAH 3.3\um\ luminosities of the $z\sim1$ galaxies correlate over $\sim$1.5 dex, supporting PAH 3.3\um\ as a star-formation-rate tracer out to $z\sim1.6$ (Figure~\ref{fig:paa_pah33}).

\item We perform an empirical characterization of the MIRI WFSS sensitivity, from both the residual background and  emission-line injection and re-extraction experiments (Figures~\ref{fig:fieldbkg} and \ref{fig:linesens_time}).
We find that single exposures beyond $\sim$1--2\,ks are no longer photon-limited and hit a systematic noise floor.
At integration $t \lesssim 4$\,ks, the achieved emission-line sensitivity is $\sim$1.8 times deeper than the prediction of the JWST ETC, whereas at the deepest available depth ($\sim$180\,ks), the ETC over-predicts the achievable sensitivity by up to $\sim$50\%.

\item Based on lessons learned from this archival census, we recommend that MIRI WFSS surveys avoid single deep integrations beyond $\sim$1--2\,ks because of this noise floor.
We prefer high-ecliptic-latitude fields, as low-latitude fields like COSMOS and UDS require $\sim$30--50\% longer integrations under the brighter zodiacal background (Figure~\ref{fig:fieldbkg}).
We also suggest designing MIRI WFSS observations through the MIRI LRS templates with a target placed on the slit, which adds a 5--10$\times$ more sensitive slit spectrum and maximizes the telescope observing efficiency (Figure~\ref{fig:mode}).
\end{enumerate}

The calibration was produced through an AI-assisted agentic workflow under human supervision, compressing an instrument-team-scale effort into much shorter timescales while keeping every step scripted and reproducible.
The data processing pipeline can be applied directly to future MIRI WFSS observations in extragalactic deep fields or Milky Way star-forming regions.

The \relname\ reference files, a worked reduction notebook (executed end-to-end on public GO-4192 exposures in GOODS-N), an example source catalog, and a MAST download script for the example data are publicly available at \url{https://github.com/fengwusun/miri_wfss} and through Zenodo at \dataset[doi:10.5281/zenodo.22803722]{https://doi.org/10.5281/zenodo.22803722}.
The \Natlas-source spectral atlas (source table and compiled 2D/1D spectra; Appendix~\ref{sec:appendix_table}) is released through Zenodo at \dataset[doi:10.5281/zenodo.22803842]{https://doi.org/10.5281/zenodo.22803842}.

\begin{acknowledgments}
J.A.-M.\ acknowledges support by grant PID2024-158856NA-I00 from the Spanish Ministry of Science and Innovation/State Agency of Research MCIN/AEI/10.13039/501100011033 and by ``ERDF A way of making Europe".
J.M.H.\ acknowledges support from the Evolving Universe Fellowship, which is made possible by a generous donation from Dr. Keiko Miwa Ross; J.M.H. also acknowledges support from JWST Program \#8544.

This work is based on observations made with the NASA/ESA/CSA James Webb Space Telescope.
The data were obtained from the Mikulski Archive for Space Telescopes at the Space Telescope Science Institute, which is operated by the Association of Universities for Research in Astronomy, Inc., under NASA contract NAS 5-03127 for \jwst.
These observations are associated with programs GO-3224, GO-4192, GO-4762, GO-6219, GO-8051, GO-8544, CAL-9265, and CAL-9505.
The specific observations analyzed can be accessed via MAST at \dataset[doi:10.17909/x4gy-3f42]{https://doi.org/10.17909/x4gy-3f42}.
FS thanks the science teams, especially the PIs of these programs, including Stacey Alberts, Seiji Fujimoto, Jed McKinney, and Andreea Petric, for designing and executing the observations that make this archival calibration and study possible.
FS specially thanks Stacey Alberts, Daniel J.\ Eisenstein, Jed McKinney and Irene Shivaei for very useful discussions regarding MIRI WFSS.

The calibration, analysis, and parts of the manuscript preparation were carried out with the assistance of Claude (Anthropic), primarily through the Claude Opus 4.8 and Claude Fable 5 models operating as an agentic coding assistant through Claude Code.
All analyses and paper writing are under the direction, review, and verification of the authors (Section~\ref{sec:cal_method}).
The authors have reviewed and verified the content of this paper and relevant calibrations, and the authors are solely responsible for all content of this paper.
\end{acknowledgments}

\begin{contribution}
F.S.\ led the data analyses and writing of the paper. 
J.A.M., E.E., R.F.A., J.M.H., X.L.\ and J.L. contributed to the data collection and analyses.
All co-authors contributed to the scientific interpretation of the results and helped to write and review the manuscript.
\end{contribution}

\facilities{JWST (MIRI, NIRCam), HST (ACS, WFC3-IR), Spitzer (IRAC)}

\software{
\texttt{astropy} \citep{astropy2022},
\texttt{numpy} \citep{harris2020},
\texttt{scipy} \citep{virtanen2020},
\texttt{matplotlib} \citep{hunter2007},
\texttt{pandeia} \citep{pandeia},
\texttt{photutils} \citep{bradley2024},
\texttt{jwst} pipeline \citep{bushouse2024},
Claude Code with Opus 4.8 and Fable 5 (Anthropic)
}



\begin{deluxetable*}{llllccc}
\tablecaption{Summary of the key \jwst/MIRI programs analyzed in this work.\label{tab:programs}}
\tablehead{
\colhead{Program} & \colhead{PI} & \colhead{Field / target} &
\colhead{Role} & \colhead{$N_{\rm exp}$\tablenotemark{a}} &
\colhead{$t_{\rm exp}$ (hr)} & \colhead{Pointings}
}
\startdata
GO-3224  & McKinney        & GOODS-N          & trace, wavelength, flux, sky, science & 54 & 12.1 & 27 \\
GO-4192  & Alberts & GOODS-N         & trace, wavelength, flux, sky, science        & 86 & 7.6  & 13 \\
GO-4762  & Fujimoto        & GOODS-N          & trace, sky, science             & 8  & 3.7  & 1  \\
GO-8544  & Helton          & GOODS-S          & trace, sky, science             & 32 & 68.1 & 1  \\
CAL-9505 & Petric          & \lmcpn     & trace, wavelength & 4 & 0.2 & 1 \\
CAL-9265 & Petric          & \hdstar          & flux (CALSPEC standard), L-flat    & 55 & 1.5  & 5  \\
GO-8051  & \'Alvarez-M\'arquez & A2744 / EGS & flat-field and sky background statistics       & 48 & 55.9 & \nodata \\
\enddata
\tablenotetext{a}{Public FULL-array P750L exposures within the listed field.
The exposure times are summed effective integration times.
GO-3224 and GO-4192 contain additional P750L exposures in other fields (22 and 102, the latter in COSMOS), which contribute only to the flat-field and sky background statistics.}
\end{deluxetable*}



\appendix
\section{Data Release of the Spectral Atlas}
\label{sec:appendix_table}

We release the MIRI WFSS spectral atlas through a Zenodo repository accompanying this publication (DOI: \dataset[10.5281/zenodo.22803842]{https://doi.org/10.5281/zenodo.22803842}).
The release contains a source table of all \Natlas\ galaxies with their key information, including JADES source ID, coordinates, spectroscopic redshift, F444W magnitude, and the number and total effective exposure time of the co-added exposures.
The IDs and coordinates follow the JADES photometric catalogs.
For each source, the release also compiles the calibrated, co-added 2D spectrum and the extracted 1D spectrum (Section~\ref{sec:atlas_extraction}).
The 1D spectra are provided with wavelengths in \micron\ and flux densities in mJy, together with propagated uncertainties and per-source metadata.
We refer the reader to Section~\ref{sec:caveats} for caveats on the use of these spectra.

\bibliographystyle{aasjournalv7}
\bibliography{00_main}

\end{document}